\documentclass[10pt, conference]{IEEEtran}

\usepackage{xcolor,colortbl}
\usepackage{hhline}
\usepackage{amsmath}
\usepackage{multirow}
\usepackage{amsfonts}
\usepackage{caption}
\usepackage{tikz}
\usepackage{booktabs}
\usepackage{pifont}
\usepackage{float}
\usepackage{enumitem}
\usepackage{tcolorbox}
\usepackage{tabularx}
\usepackage[flushleft]{threeparttable}
\usepackage[hidelinks]{hyperref}
\usepackage[T1]{fontenc}
\usepackage[titlenumbered,ruled,noline,linesnumbered]{algorithm2e}

\usetikzlibrary{calc}
\usetikzlibrary{patterns}
\usetikzlibrary{shapes.multipart, arrows, positioning,shapes.geometric}
\usetikzlibrary{decorations.pathreplacing}
\usetikzlibrary{calc,shapes,decorations,decorations.shapes,positioning,shadows,arrows,decorations.markings,backgrounds,fit}

\makeatletter
\newcommand*{\rom}[1]{\expandafter\@slowromancap\romannumeral #1@}
\makeatother
\renewcommand{\vec}[1]{\mathbf{#1}}

\begin{document} 

\title{Adversarial Neural Network Inversion via Auxiliary Knowledge Alignment}

\author{\IEEEauthorblockN{Ziqi Yang,
Ee-Chien Chang, Zhenkai Liang}
\IEEEauthorblockA{School of Computing, National University of Singapore\\
\{yangziqi, changec, liangzk\}@comp.nus.edu.sg}}

\maketitle

\begin{abstract}
The rise of deep learning technique has raised new privacy concerns about the training data and test data. 
In this work, we investigate the model inversion problem in the adversarial settings, where the adversary aims at inferring information about the target model's training data and test data from the model's prediction values. 
We develop a solution to train a second neural network that acts as the inverse of the target model to perform the inversion.
The inversion model can be trained with black-box accesses to the target model.
We propose two main techniques towards training the inversion model in the adversarial settings.
First, we leverage the adversary's background knowledge to compose an auxiliary set to train the inversion model, which does not require access to the original training data.
Second, we design a truncation-based technique to align the inversion model to enable effective inversion of the target model from partial predictions that the adversary obtains on victim user's data.
We systematically evaluate our inversion approach in various machine learning tasks and model architectures on multiple image datasets.
Our experimental results show that even with no full knowledge about the target model's training data, and with only partial prediction values, our inversion approach is still able to perform accurate inversion of the target model, and outperform previous approaches.

\end{abstract}

\section{Introduction}

Machine learning (ML) models, especially deep neural networks, are
becoming ubiquitous, powering an extremely wide variety of
applications. The mass adoption of machine learning technology has
increased the capacity of software systems in large amounts of complex
data, enabling a wide range of applications. For example, facial recognition APIs,
such as Apple's face-tracking APIs in the ARKit~\cite{apple_face},
provide scores of facial attributes, including emotion and
eye-openness levels. There are also online services that evaluate
users' face beauty and age~\cite{how-old, beauty-ai, xiaoice}.  Users
often share their face-beauty scores on social media for fun. The
prediction scores are definitely linked to the input face data, but it
is not apparent to what level of accuracy the original data can be
recovered. Privacy concerns arise in such applications.

\begin{figure}[t]
\begin{center}
\includegraphics[width=\linewidth]{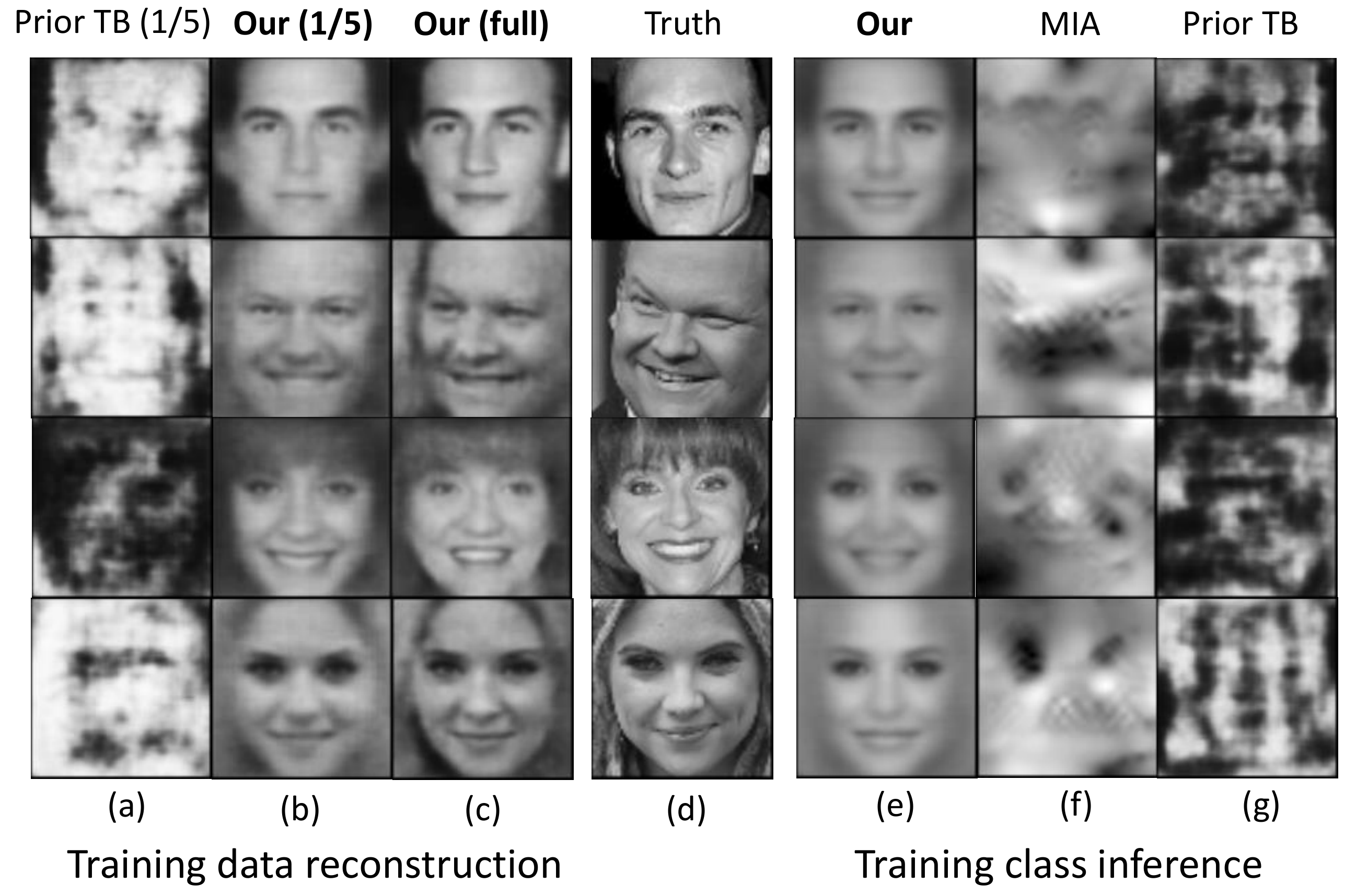}
\caption{Training data reconstruction and training class inference of
  our and previous approaches against a facial recognition classifier.
  Column (a) and (b) compare our approach with a known
  \underline{t}raining-\underline{b}ased (TB) method when only 1/5
  prediction results are available. Column (c) shows our result of
  inverting full prediction vectors by training the inversion model on
  auxiliary samples. Column (e)-(g) compare our approach with MIA and
  the known prior TB method. Note that we use auxiliary samples for our
  approach in all experiments.}
\label{fig:killer_figure}
\end{center}
\end{figure}

There have been many research efforts in {\em model inversion}, which
aims to obtain information about the training data from the
model's predictions. They are largely divided into two classes of approaches. The
first class inverts a model by making use of gradient-based
optimization in the data
space~\cite{model_inversion,inverse_local_global, inverse_mlp,
  inverse_linear_nonlinear, inverse_feedforward,
  inverse_measurement_control, understand_DNN_invert}.
  We call this class of approach
  \textit{optimization-based} approach.
For example, model inversion attack (MIA)~\cite{model_inversion} was
proposed to infer training classes against neural
networks by generating a representative sample for the target class. It casts the inversion task as an optimization problem to
find the ``optimal'' data for a given class.  MIA works for simple
networks but is shown to be ineffective against complex neural networks such
as convolutional neural network (CNN)~\cite{sok_ml_security,
  membership, ccs_gan}. Our experiment also obtains similar result as shown in Figure~\ref{fig:killer_figure}
column (f).  
This is mainly because the optimization objective of optimization-based approach does not really capture the semantics of the data space (see Section~\ref{sec:model_inversion}).
The second class of approach~\cite{nn_recognition, invert_visual_cnn,
  nips_inversion, inverse_autoregressive} invert a model by learning a
second model that acts as the inverse of the original one.
We call this class of approach
\textit{training-based} approach.
They target
at reconstructing images from their computer vision features including
activations in each layer of a neural network. Therefore, to maximally
reconstruct the images, they train the second model using \textit{full}
prediction vectors on the \textit{same} training data of the
classifier.

In this work, we focus on the adversarial scenario, where an adversary
is given the classifier and the prediction result, and aims to make
sense of the input data and/or the semantics of the
classification. Such problem is also known as the {\em inversion
  attack}~\cite{model_inversion, fredrikson2014privacy} wherein
information about the input data could be inferred from its prediction
values.  There are two adversarial inversion settings, namely
\textit{data reconstruction} and \textit{training class inference}.

{\em Data reconstruction:} In this attack, the adversary is asked to
reconstruct unknown data given the classifier's prediction vector on
it, which is exactly the inversion of the classifier. 
For example, in a facial recognition classifier that
outputs the person's identity in a facial image,
the attacker's goal is to reconstruct the
facial image of a person. 

{\em Training class inference:} This kind of attack aims at recovering
a semantically meaningful data for each training class of a trained
classifier. In the above-mentioned
facial recognition example, the attack goal is to
recover a recognizable facial image of an arbitrary person (class) in
the training data.  This attack can be achieved by inverting the classifier
to generate a representative sample which is classified as the wanted
class~\cite{model_inversion}.

However, inversion in adversarial settings is different from existing
model inversion settings. Specifically, in adversarial settings, the
attacker has no knowledge about the classifier's training data.
Thus, previously known training-based methods that use the same training data to train the ``inversion'' model cannot be directly applied. 
Existing optimization-based approaches require white-box accesses to the classifier to compute the gradients, and therefore they cannot work if the classifier is a blackbox.
More importantly, the adversary
might obtain only partial prediction results on victim data.  For
example, the ImageNet dataset~\cite{imagenet} has over 20,000 classes.
Since the majority values in the prediction vector are small, it is more
practical to release only the top 5 to 10 predicted
classes~\cite{train_imagenet}.  Victim users might also post partial
predicted scores to the social media.  The partial predictions largely
limit the reconstruction ability of the inversion model.  For example,
Figure~\ref{fig:killer_figure} column (a) shows the result of
inverting a ``truncated'' prediction vector (i.e., keeping the 1/5
largest values while masking the rest to 0).  The poor result is probably due to
overfitting in the inversion model. Hence, although reconstruction
is accurate on the full prediction vector, a slight deviation such as
truncation in the prediction vector will lead to a dramatically
different result.

In this paper, we formulate the adversarial settings, and propose an
effective approach to invert neural networks in such settings.
We adopt the above-mentioned training-based approach. That is, a
second neural network (referred to as \textit{inversion model}) is
trained to perform the inversion. In contrast to previous work,
instead of using the same distribution of the training data, in our
setting, the training set (of the inversion model) is drawn from a
more generic data distribution based on the background knowledge which is arguably much easier for the
adversary to obtain.  For instance, against a facial recognition
classifier, the adversary could randomly crawl facial images from the
Internet to compose an auxiliary set without knowing the exact
training data (distribution).
Figure~\ref{fig:killer_figure} column (c) shows the reconstruction
result of the inversion model trained on auxiliary samples, which clearly outperforms the result in column (a). 
We believe that this is because the auxiliary samples 
retain generic facial features (e.g., face edges, eyes and nose)
and such information is sufficient to regularize the originally
ill-posed inversion problem.  Interestingly, our experiments show that
even in cases where the actual training data is available, augmenting
the training data with generic samples could improve inversion
accuracy.

To make the inversion model also work when the adversary obtains only
partial prediction result on victim user's data, we propose a truncation
method of training the inversion model.  The main idea is to feed the
inversion model with truncated predictions (on the auxiliary
samples) during training. Such truncation forces the inversion model to
maximally reconstruct the samples based on the truncated predictions,
and thus align the inversion model to the truncated
values. Furthermore, truncation helps to reduce overfitting in a similar way of feature selection~\cite{feature_selection}.
Figure~\ref{fig:killer_figure} column (b) shows the result of
inverting a partial prediction vector (i.e., keeping the 1/5 largest
values while masking the rest to 0) using our truncation technique.

It turns out that the truncation technique is effective in our second
problem on training class inference.  To infer the training classes,
we truncate the classifier's predictions on auxiliary samples to
one-hot vectors. After the training is complete,
it can produce a representative sample for each training class by
taking one-hot vectors of the classes as input.
Figure~\ref{fig:killer_figure} column (e) shows the result of our
training class inference. It significantly outperforms MIA (column
(f)) and previous training-based approach (column (g)).

Depending on the adversary's role, we design two ways
of training the inversion model.
First, if the adversary is a user with no knowledge about the training
data and has black-box access to the classifier (i.e., only prediction
results are available to him), we construct the inversion model by
training a separate model from sketch. Second, if the adversary is the developer who trains the
classifier, the inversion model can be trained jointly with the
classifier on the same training data, which derives a more precise
inversion model.  
The objective is to force the classifier's predictions to also
preserve essential information which helps the inversion model to
reconstruct data. To this end, we use the reconstruction loss of
the inversion model to regularize the training of the
classifier.  

We evaluate our approaches for various machine learning tasks and
model architectures on multiple image datasets. The results show that
our attack can precisely reconstruct data from partial predictions
without knowing the training data distribution, and also outperforms previous approaches in training class inference.  
This work highlights that the rich information hidden in the model's prediction can be extracted, even if the adversary only has access to limited information.
We hope for this work to contribute to strengthening user's awareness of handling derived data.

\textbf{Contributions.} In summary, we make the following
contributions in this paper.
\begin{itemize}

\item We observe that even with partial knowledge of the training set, it is possible to exploit such background knowledge to regularize the originally ill-posed inversion problem.

\item We formulate a practical scenario where the adversary obtains
  only partial prediction results, and propose a truncation method that aligns the inversion model to the space of the truncated predictions.

\item Our inversion method can be adopted for training class
  inference attack. Experimental results show that our method
  outperforms existing work on training class inference.

\item We observe that even if the classifier's training set is available
  to the adversary, augmenting it with generic samples
  could improve inversion accuracy of the inversion model.

\end{itemize}

\section{Background}

\subsection{Machine Learning}
\label{sec:ml_background}
In this paper, we focus on supervised learning, more specifically, on training classification models (classifiers) using neural networks~\cite{DL_origin}. The model is used to give predictions to input data. The lifecycle of a model typically consists of \textit{training} phase (where the model is created) and \textit{inference} phase (where the model is released for use).

\textit{\underline{Machine learning models.}} 
A machine learning classifier encodes a general hypothesis function $F_w$ (with parameters $w$) which is learned from a training dataset with the goal of making predictions on unseen data.  
The input of the function $F_w$ is a data point $\vec{x}\in\mathbb{R}^d$ drawn from a data distribution $p_x(\vec{x})$
in a $d$-dimensional space $\mathcal{X}$, where each dimension represents one attribute (feature) of the data point. The output of $F_w$ is a predicted point $F_w(\vec{x})$ in a $k$-dimensional space $\mathcal{Y}$, where each dimension corresponds to a predefined class.
The learning objective is to find the relation between each input data and the class as a function $F_w:\mathcal{X}\mapsto\mathcal{Y}$.
A neural network (deep learning model) consists of multiple connected layers of basic non-linear activation functions (neurons) whose connections are weighted by the model parameter $w$, with a normalized exponential function $\mathsf{softmax}(\vec{z})_i = \frac{\exp(z_i)}{\sum_j \exp(z_j)}$ added to the activation signals (referred to as \textit{logits}) of the last layer. This function converts arbitrary values into a vector of real values in $[0,1]$ that sum up to $1$.  Thus, the output could be interpreted as the probability that the input falls into each class.  

\textit{\underline{Training phase.}} 
Let $\vec{x}$ represent the data drawn from the underlying data distribution $p_x(\vec{x})$, and $\vec{y}$ be the vectorized class of $\vec{x}$.
The training goal is to find a function $F_w$ to well approximate the mapping between every data point $(\vec{x},\vec{y})$ in space $\mathcal{X}\times\mathcal{Y}$. To this end, we use a loss function $\mathcal{L}(F_w(\vec{x}),\vec{y})$ to measure the difference between the class $\vec{y}$ and the classifier's prediction $F_w(\vec{x})$. Formally, the training objective is to find a function $F_w$ which minimizes the expected loss.
\begin{equation}
L(F_w) = \mathbb{E}_{\vec{x}\sim p_x} [\mathcal{L}(F_w(\vec{x}), \vec{y})]
\end{equation}

The actual probability function $p_x(\vec{x})$ is intractable to accurately represent, but in practice, we can estimate it using samples drawn from it. These samples compose the training set $D\subset\mathcal{X}$. 
We predefine a class $\vec{y}$ for each data $\vec{x}\in D$ as supervision in the training process.
Hence, we can train the model to minimize the empirical loss over the training set $D$.
\begin{equation}
\label{classification_loss}
L_D(F_w) = \frac{1}{|D|} \sum_{\vec{x}\in D} \mathcal{L} (F_w(\vec{x}),\vec{y})
\end{equation}

Nonetheless, this objective could lead to an overfitted model which attains a very low prediction error on its training data, but fails to generalize well for unseen data drawn from $p_x(\vec{x})$. A regularization term $R(F_w)$ is usually added to the loss $L_D(F_w)$ to prevent $F_w$ from overfitting on the training data. In summary, the training process of a classifier is to find a model $F_w$ that minimizes the following objective:
\begin{equation}
\label{regualarized_loss}
\mathcal{C}(F_w)=L_D(F_w)+\lambda R(F_w)
\end{equation}
where the regularization factor $\lambda$ controls the balance between the classification function and the regularization function.

Algorithms used for solving this optimization problem are variants of the gradient descent algorithm~\cite{gradient_descent}.  Stochastic gradient descent (SGD)~\cite{zhang2004solving} is a very efficient method that updates the parameters by gradually computing the average gradient on small randomly selected subsets (mini-batches) of the training data.

\textit{\underline{Inference phase.}}
In the inference phase (or called testing phase), the model $F_w$ is used to classify unseen data.
Specifically, function $F_w$ takes any data $\vec{x}$ drawn from the same data distribution $p_x(\vec{x})$ as input, and outputs a prediction vector $F_w(\vec{x})=(F_w(\vec{x})_1,...,F_w(\vec{x})_k)$, where $F_w(\vec{x})_i$ is the probability of the data $\vec{x}$ belonging to class $i$ and $\sum_i{F_w(\vec{x})_i}=1$.

\subsection{Model Inversion}
\label{sec:model_inversion}

Our approach is related to many previous work on inverting neural networks from machine learning and computer vision communities.
Inverting a neural network helps in understanding and interpreting the model's behavior and feature representations.
For example, a typical inversion problem in computer vision is to reconstruct an image $\vec{x}$ from its computer vision features such as HOG~\cite{hog} and SIFT~\cite{sift}, or from the activations in each layer of the network including the classifier's prediction $F_w(\vec{x})$ on it.
In general, these inversion approaches fall into two categories: optimization-based inversion~\cite{model_inversion,inverse_local_global, inverse_mlp, inverse_linear_nonlinear, inverse_feedforward, inverse_measurement_control, understand_DNN_invert} and, most similar to our approach, training-based inversion~\cite{nn_recognition, invert_visual_cnn, nips_inversion, inverse_autoregressive}.

\textit{\underline{Optimization-based inversion.}} 
The basic idea of this branch of work is to apply gradient-based optimization in the input space $\mathcal{X}$ to find an image $\hat{\vec{x}}$ whose prediction approximates a given $F_w(\vec{x})$. To this end, the image $\hat{\vec{x}}$ should minimize some loss function between $F_w(\vec{x})$ and $F_w(\hat{\vec{x}})$.
However, inverting the prediction of a neural network is actually a difficult ill-posed problem~\cite{nips_inversion}. 
The optimization process tends to produce images that do not really resemble natural images especially for a large neural network~\cite{yosinski2015understanding}. 
To mitigate this issue, a series of studies have investigated using a natural image prior $P(\hat{\vec{x}})$ to regularize the optimization. The prior defines some statistics of the image. Formally, the inversion is to find an $\hat{\vec{x}}$ which minimizes the following loss function.
\begin{equation}
\mathcal{O}(\hat{\vec{x}}) = \mathcal{L}(F_w(\hat{\vec{x}}),F_w(\vec{x})) + P(\hat{\vec{x}})
\end{equation}
where $\mathcal{L}$ is some distance metric such as L2 distance.
A special case of such inversion is to generate a representative image for some class $y$, by replacing $F_w(\vec{x})$ with vectorized $y$ in this equation~\cite{model_inversion,saliency_map}. 
Optimization-based inversion requires white-box access to the model to compute the gradients.

Various kinds of image priors have been investigated in the literature. 
For instances, a common prior is $\alpha$-norm $P_\alpha(\vec{x})=||\vec{x}||_\alpha^\alpha$, which encourages the recovered image to have a small norm. Simonyan et al.~\cite{saliency_map} use L2 norm while Mahendran and Vedaldi~\cite{understand_DNN_invert} demonstrate that a relatively large $\alpha$ produces better result. They choose L6 norm in their experiments. 
Mahendran and Vedaldi~\cite{understand_DNN_invert} also investigate using total variation (TV) $P_{V^\beta}(\vec{x})$ for the image prior, which can encourage images to have piece-wise constant patches. It is defined in the following.
\begin{equation}
\label{tv_loss}
P_{V^\beta}(\vec{x})=\sum_{i,j}\left(\left(x_{i,j+1}-x_{i,j}\right)^2+\left(x_{i+1,j}-x_{i,j}\right)^2\right)^{\beta/2}
\end{equation}
Besides, randomly shifting the input image before feeding it to the network is also used to regularize the optimization in recent work~\cite{visulize_cnn_pre_image}.
In addition, Yosinski et al.~\cite{yosinski2015understanding} investigate a combination of three other priors. Gaussian blur is used to penalize high frequency information in the image. They clip pixels with small norms to preserve only the main object in the image. Besides, they also clip pixels with small contribution to the activation. The contribution measures how much the activation increases or decreases when setting the pixel to 0.
Fredrikson et al.~\cite{model_inversion} adopt denoising and sharpening filters as the prior in the model inversion attack (MIA). MIA targets at generating a representative image for training classes.

However, the simple hand-designed prior $P$ in these approaches is limited.
It cannot really capture the semantic information in the training data space, so the reconstruction quality especially against large networks is not satisfactory. Furthermore, this approach involves optimization at the test time because it requires computing gradients which makes it relatively slow (e.g., 6s per image on a GPU~\cite{understand_DNN_invert}).

\textit{\underline{Training-based inversion.}} This kind of inversion trains another neural network $G_\theta$ (referred to as \textit{inversion model} in this paper) to invert the original one $F_w$. Specifically, given the \textit{same} training set of images and their predictions $(F_w(\vec{x}),\vec{x})$, it learns a second neural network $G_\theta$ from sketch to well approximate the mapping between predictions and images (i.e., the inverse mapping of $F_w$). 
The inversion model $G_\theta$ takes the prediction $F_w(\vec{x})$ as input and outputs an image.
Formally, this kind of inversion is to find a model $g$ which minimizes the following objective.
\begin{equation}
\label{eq:recon_loss}
\mathcal{C}(G_\theta)=\mathbb{E}_{\vec{x}\sim p_x}[\mathcal{R}(G_\theta(F_w(\vec{x})), \vec{x})]
\end{equation}
where $\mathcal{R}$ is the image reconstruction loss such as L2 loss adopted in work~\cite{invert_visual_cnn}.
Dosovitskiy and Brox~\cite{nips_inversion} investigate using two additional loss items to regularize the training of the inversion model: loss in feature space and adversarial loss.
Specifically, the loss in feature space encourages the reconstructed image to preserve perceptually important image features. The features, for example, can be the activations in some layer of the inversion model.
The adversarial loss is added to keep the reconstructions realistic. Following the spirit of Generative Adversarial Networks (GANs)~\cite{goodfellow2014generative}, it uses a discriminator $D_\phi$ to discriminate the reconstructed images from real ones, while the inversion model is trained to ``fool'' the discriminator into classifying the reconstructed images as real. The adversarial loss is defined in the following.
\begin{align}
\label{adv_loss}
\begin{split}
\mathcal{A}(G_\theta,D_\phi)= & \mathbb{E}_{\vec{x}\sim p_x}[\log(D_\phi(\vec{x}))] \\
&+\mathbb{E}_{\vec{x}\sim p_x}[\log(1-D_\phi(G_\theta(F_w(\vec{x}))))]
\end{split}
\end{align}

In contrast to optimization-based inversion, training-based inversion is only costly during training the inversion model which is one-time effort. Reconstruction from a given prediction requires just one single forward pass through the network (e.g., 5ms per image on a GPU~\cite{invert_visual_cnn}).

\section{Adversarial Model Inversion}

The adversary could be either the user of a blackbox classifier $F_w$,  or the developer of  $F_w$. The adversary's capabilities and goals  differ depending on the role and we consider three  scenarios: (1) A curious user who intends to reconstruct the victim's input data from the victim's truncated predication vector; (2)  A curious user who intends to  infer  $F_w$'s functionality;  (3) A malicious developer who intends to build a $F_w$, which subsequently  could  help in  reconstructing the victims' input from their truncated predication vectors.

\subsection{(Scenario 1) Data reconstruction with blackbox classifier}

In this scenario,  the adversary is a curious user who has  black-box accesses to a classifier $F_w$. That is, the adversary can adaptively feed input to $F_w$ and obtain the output.  The adversary does not know  the classifier's training data (distribution), architecture and parameters.  However, the adversary has some background knowledge on $F_w$. Specifically, the adversary  know the following:
\begin{itemize}
\item  Although the adversary does not know the actual training data that is used to train $F_w$, the adversary can draw samples from a more ``generic'' distribution $p_a$ of the training data. For instance, suppose $F_w$  is a face recognition classifier trained on faces of a few individuals,  although  the adversary does not know the faces of those individuals,    the adversary knows that the training data are facial images and thus can draw many samples from a large pool of facial images.  Intuitively, a distribution $p_a$ is more generic than the original one if $p_a$ is the  distribution after some dimension reductions are applied on the original. 

\item  The adversary knows the input format of $F_w$, since he knows the distribution $p_a$. The adversary also knows the output format of $F_w$. That is, the adversary know the dimension of the predication vector. This assumption is reasonable because even though $F_w$ may return selected prediction values,  the adversary can still estimate the dimension of the prediction vector. For example, he can query $F_w$ with a set of input data and collect distinct classes in the returned predictions as the dimension. 
\end{itemize}

The classifier $F_w$ is also used  by many other benign users. We assume that the adversary has the capability to obtain ${\vec f}$, a  $m$-{\em truncated} predication vector of $F_w({\vec x})$ where ${\vec x}$ is the input of a victim user, and $m$ is a predefined parameter determined by the victim. 
Given a prediction vector ${\vec g}$, we say that ${\vec f}$ is $m$-truncated, denoted as $\mathsf{trunc}_m({\vec g})$,  when all $m$ largest values of  ${\vec g}$ remain in ${\vec f}$, while the rest are truncated  to zeros.  For instance,
$$\mathsf{trunc}_2(  ( 0.6,  0.05, 0.06, 0.2, 0.09 ) ) = ( 0.6, 0, 0, 0.2, 0 ).$$

Now, given ${\vec f}$, black-box access to $F_w$, and samples from distribution $p_a$, the adversary wants to find a most probable data from the distribution $p_a$ such  that $\mathsf{trunc}(F_w(\vec{x}))= {\vec f}$.  That is, ideally, the adversary wants to find a  $\hat{\vec{x}}$ that satisfies the following:
\begin{align}
\begin{split}
\hat{\vec{x}}&=\arg\max_{\vec{x} \in X_{\vec f}} p_a(\vec{x}) \\
\text{subject to} \quad  X_{{\vec f}}&=\{\vec{x}\in\mathcal{X} \mid \mathsf{trunc}(F_w(\vec{x}))= {\vec f}\}
\end{split}
\end{align}
Let us call the problem of obtaining  $\hat{\vec{x}}$ from $F_w$, ${\vec f}$ and $p_a$ the {\em data reconstruction} problem.

\subsection{(Scenario 2) Training class inference}
Same as in scenario 1,  in this scenario,  the adversary is a curious user who has  black-box accesses to a classifier $F_w$, and knows samples from a  generic distribution $p_a$.

Instead of data reconstruction, the adversary wants to find a representative data of the training class.    
Given black-box access to a classifier $F_w$, and a target class $y$,  the adversary wants to  find a data $\hat{\vec{x}}$ that satisfies the following.
\begin{align}
\begin{split}
 \hat{\vec{x}}&=\arg\max_{\vec{x} \in X_{y}} p_a(\vec{x}) \\
\text{subject to} \quad X_{y}&=\{\vec{x}\in\mathcal{X} \mid 
F_w(\vec{x})_y \text{ is high}\}
\end{split}
\end{align}
where $F_w(\vec{x})_y$ is the confidence that $\vec{x}$ is classified by $F_w$ as class $y$.

\subsection{(Scenario 3) Joint  classifier and model inversion}

We also consider the scenario where the adversary is a malicious developer who trains the classifier  $F_w$ and sells/distributes it to users. Different from the previous two scenarios,  here, the adversary  has full knowledge about the classifier's training data, architecture and parameters, and  has the freedom to decide what  $F_w$ would be.    We assume that after $F_w$ is released, the adversary is able to obtain truncated predication from the users, and the adversary wants to reconstruct the users' input. 

Hence, in this scenario, the adversary goal is to  train a classifier  $F_w$   that meets the accuracy requirement (w.r.t. the original classifier's task), while improving quality of data reconstruction.

\subsection{Applying Prior Work in Adversarial Settings}

Let us highlight the difference in the adversary's capabilities between our adversarial settings and previous inversion settings. 

First, in our adversarial settings, when the adversary is a user, if he has only black-box accesses to the classifier, existing optimization-based inversion approaches including MIA do not work because they require white-box access to compute the gradients. Besides, many studies have shown that MIA against large neural networks tends to produce semantically meaningless images that do not even resemble natural images~\cite{membership, ccs_gan}. 
Our experimental result, as shown in Figure~\ref{fig:killer_figure} column (f), also reaches the same conclusion.
Second, the adversary as a user has no knowledge about the classifier's training data (distribution), which makes previous training-based approaches impossible to train the inversion model on the same training data.
Third, the adversary, as either a user or developer, might obtain truncated prediction results. This makes previous inversion models that are trained on full predictions ineffective in reconstructing data from partial predictions, as shown in Figure~\ref{fig:killer_figure} column (a) and (g).

\section{Approach}

\begin{figure}[t]
\begin{center}
\includegraphics[width=\linewidth]{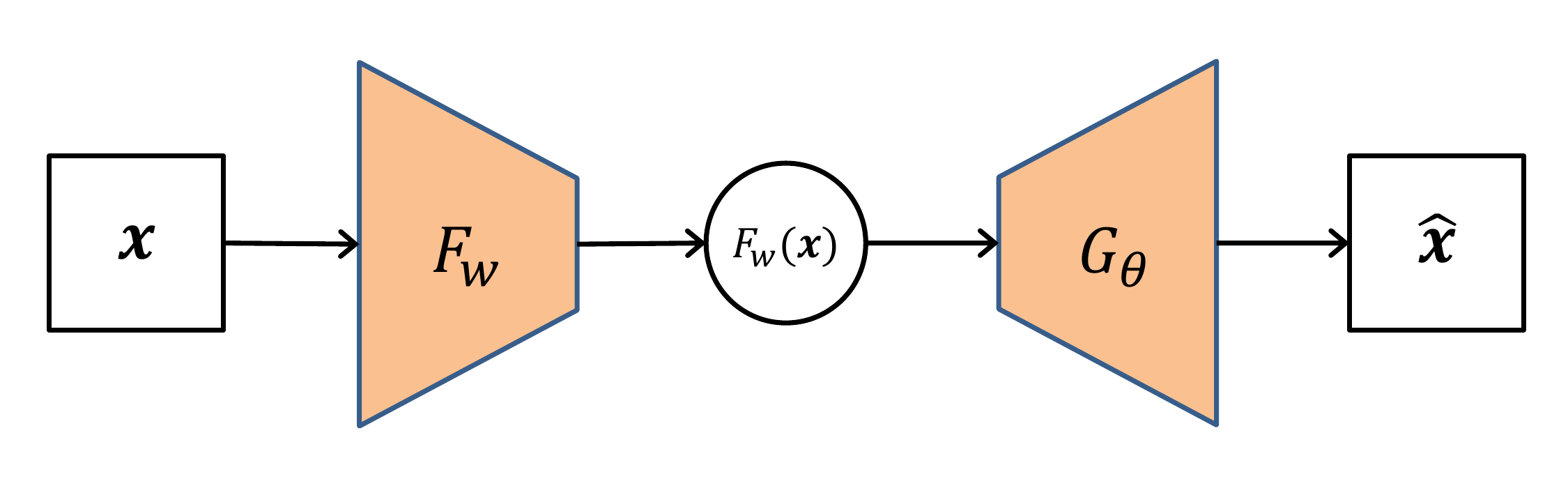}
\caption{Framework of training-based inversion approach. The classifier $F_w$ takes data $\vec{x}$ as input and produces a prediction vector $F_w(\vec{x})$. The inversion model $G_\theta$ outputs the reconstructed data $\hat{\vec{x}}$ with the prediction as input.}
\label{fig:autoencoder}
\end{center}
\end{figure}

Our approach adopts previously mentioned training-based strategy to invert the classifier. The overall framework is shown in Figure~\ref{fig:autoencoder}. 
Unlike the optimization-based approaches that invert a given prediction vector directly from $F_w$, here, an inversion model $G_\theta$ is first trained, which later takes the given prediction vector as input and outputs the reconstructed sample.
This is similar to autoencoder~\cite{autoencoder} where $F_w$ is the ``encoder'', $G_\theta$ is the ``decoder'', and the prediction is the latent space.

There are three aspects that are different from autoencoder, which we will elaborate in this section. (1) When $F_w$ is fixed (Scenario 1\&2), we need training data to train $G_\theta$ as we highlight earlier that the adversary does not have the data that trains $F_w$. Section~\ref{sec:aux_set} describes how we obtain the training set for $G_\theta$. (2) Since our adversary only obtains truncated prediction, we need a method to ``realign'' the latent space. Section~\ref{sec:truncation} gives the proposed truncation method. (3) In our Scenario 3, the $F_w$ is not fixed and the adversary can decide $F_w$. In this case, this is a joint classifier and inversion model problem. Section~\ref{sec:joint_train} gives our training method.

\subsection{$G_\theta$'s Training Data: Auxiliary Set}
\label{sec:aux_set}

The first important component in the construction of $G_\theta$ is its training set, which is referred to as \textit{auxiliary set} in the rest of the paper.
The auxiliary set should contain sufficient semantic information to regularize the ill-posed inversion problem.

We compose the auxiliary set by drawing samples from a more generic data distribution $p_a$ than the original training data distribution $p_x$. For example, against a facial recognition classifier of 1,000 individuals, the auxiliary samples could be composed by collecting public facial images of random individuals from the Internet.
These auxiliary samples still retain general facial features such as the face edges and locations of eyes and nose, which are shared semantic features of the original training data. 
We believe that such shared features provide sufficient information to regularize the originally ill-posed inversion task.
To further improve the reconstruction quality of $G_\theta$, we can specially choose the auxiliary set to better align the inversion model. For example, against a facial recognition classifier, we choose the dataset with mostly frontal faces as the auxiliary set, so as to align the inversion model to frontal faces.

Our experimental results demonstrate the effectiveness of sampling auxiliary set from a more generic data distribution, as shown in Section~\ref{sec:effect_of_aux}. The inversion model can precisely reconstruct the training data points even if it never sees the training classes during the construction.

\subsection{Truncation Method for Model Inversion}
\label{sec:truncation}

\def\layersep{1.5cm}
\def\modelsep{2cm}
\def\nodesep{1cm}

\tikzset{neuron/.style={circle,thick,fill=white,draw,minimum size=15pt,inner sep=0pt},
    hoz/.style={rotate=-90}, 
    prediction/.style={
            draw,
            rectangle split,
            rectangle split parts=10,
            rectangle split empty part height=0.2cm,
            thick,fill=white,minimum height=4cm,minimum width=0.6cm
        },
     filtered/.style={
                 draw,
                 rectangle split,
                 rectangle split parts=6,
                 rectangle split empty part height=0.2cm,
                 thick,fill=white,minimum height=4cm,minimum width=0.6cm
             },
    }   

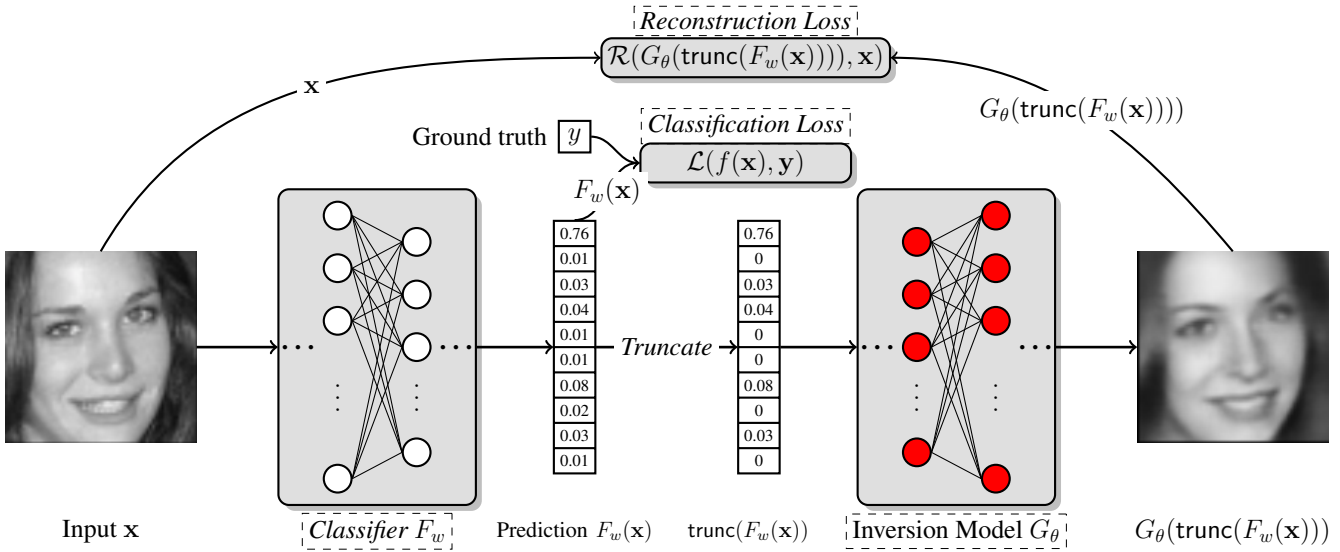
\begin{figure*}[t!]
\centering      
\begin{tikzpicture}[-,draw=black, node distance=\layersep,transform shape,rotate=0,scale=0.7]  

\tikzstyle{every node}=[font=\Large]

\node[draw,thick,drop shadow,fill=gray!25,minimum height=6*\nodesep,minimum width=2.5*\layersep,rounded corners] (classifier) at (2.5*\layersep, -3*\nodesep) {};
\node[draw,thick,drop shadow,fill=gray!25,minimum height=6*\nodesep,minimum width=2.5*\layersep,rounded corners] (inverse) at (8.5*\layersep+\modelsep, -3*\nodesep) {};

\node[inner sep=0pt] (input) at (-1*\layersep,-3*\nodesep)  {\includegraphics[width=.2\textwidth]{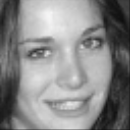}};
\node[inner sep=0pt] (recovered) at (12*\layersep+\modelsep,-3*\nodesep)  {\includegraphics[width=.2\textwidth]{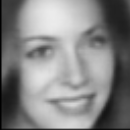}};

\node[prediction, font=\small](prediction) at (5*\layersep, -3*\nodesep) {0.76 \nodepart{two} 0.01\nodepart{three} 0.03\nodepart{four} 0.04\nodepart{five} 0.01\nodepart{six} 0.01\nodepart{seven} 0.08\nodepart{eight} 0.02\nodepart{nine} 0.03\nodepart{ten} 0.01};
\node[prediction, font=\small](filtered) at (6*\layersep+\modelsep, -3*\nodesep) {0.76 \nodepart{two} 0\nodepart{three} 0.03\nodepart{four} 0.04\nodepart{five} 0\nodepart{six} 0\nodepart{seven} 0.08\nodepart{eight} 0\nodepart{nine} 0.03\nodepart{ten} 0};

\path[->,thick,line width=1pt,rotate=90] (input) edge (classifier);

\path[->,thick,line width=1pt,rotate=90] (classifier) edge (prediction);

\path[->,thick,line width=1pt,rotate=90] (prediction) edge node [fill=white] {\rotatebox{270}{\em Truncate}} (filtered);

\path[->,thick,line width=1pt,rotate=90] (filtered) edge (inverse);

\path[->,thick,line width=1pt,rotate=90] (inverse) edge (recovered);

\node (a) at (-1*\layersep, -6.5*\nodesep) {Input $\vec{x}$};
\node[draw,dashed, line width=0.2pt] (b) at (2.5*\layersep, -6.5*\nodesep) {\em Classifier $F_w$};
\node[draw,dashed, line width=0.2pt] (c) at (8.5*\layersep+\modelsep, -6.5*\nodesep) {Inversion Model $G_\theta$};
\node (d) at (5*\layersep, -6.5*\nodesep) {\large Prediction $F_w(\vec{x})$};
\node (e) at (6*\layersep+0.9*\modelsep, -6.5*\nodesep) {\large $\mathsf{trunc}(F_w(\vec{x}))$};
\node (f) at (12*\layersep+\modelsep, -6.5*\nodesep) {$G_\theta(\mathsf{trunc}(F_w(\vec{x})))$};

\node[draw,thick,drop shadow,fill=gray!25,minimum height=0.5cm,minimum width=4cm,rounded corners] (loss) at (6.5*\layersep+0.5*\modelsep, 2.5) {$\mathcal{R} (G_\theta(\mathsf{trunc}(F_w(\vec{x})))),\vec{x})$};
\node[draw,dashed, line width=0.2pt] (c) at (6.5*\layersep+0.5*\modelsep, 3.2) {\em Reconstruction Loss};

\path[->,thick,line width=0.8pt, out=60, in=180] (input.north) edge node [fill=white] {$\vec{x}$} (loss.west);
\path[->,thick,line width=0.8pt, out=130, in=0] (recovered.north) edge node [fill=white] {$G_\theta(\mathsf{trunc}(F_w(\vec{x}))))$} (loss.east);

\node[draw,thick,drop shadow,fill=gray!25,minimum height=0.5cm,minimum width=4cm,rounded corners] (loss2) at (6.5*\layersep+0.5*\modelsep, 0.5) {$\mathcal{L} (f(\vec{x}),\vec{y})$};
\node[draw,dashed, line width=0.2pt] (c) at (6.5*\layersep+0.5*\modelsep, 1.2) {\em Classification Loss};
\node[draw,thick,fill=white,minimum height=0.6cm,minimum width=0.6cm] (label) at (5*\layersep, 1*\nodesep) {$y$};
\node (description) at (3.8*\layersep, 1*\nodesep) {Ground truth};
\path[->,thick,line width=0.8pt, out=0, in=180] (label.east) edge  (loss2.west);
\path[->,thick,line width=0.8pt, out=0, in=180,font=\small] (prediction.north) edge node [fill=white] {$F_w(\vec{x})$} (loss2.west);

\node[hoz,rotate=90] (I-other) at (1.5*\layersep,-3*\nodesep) {\textbf{. . .}};
\node[hoz,rotate=90] (H-other) at (3.5*\layersep,-3*\nodesep) {\textbf{. . .}};
\node[hoz,rotate=90] (I2-other) at (7.5*\layersep+\modelsep,-3*\nodesep) {\textbf{. . .}};
\node[hoz,rotate=90] (I2-other) at (9.5*\layersep+\modelsep,-3*\nodesep) {\textbf{. . .}};

\foreach \name / \y in {1/1*\nodesep,2/2*\nodesep,3/3*\nodesep,6/6*\nodesep}
	\path node [neuron, hoz] (I-\name) at (2*\layersep,-\y+0.5*\nodesep) {};
\node[hoz] (I-4) at (2*\layersep,-4*\nodesep) {$\dots$};

\foreach \name / \y in {1/1*\nodesep,2/2*\nodesep,3/3*\nodesep,5/5*\nodesep}
	\node[neuron, hoz] (H-\name) at (3*\layersep,-\y) {};
\node[hoz] (H-4) at (3*\layersep,-4*\nodesep) {$\dots$};

    \foreach \source in {1,2,3,6}
        \foreach \dest in {1,2,3,5}
            \path (I-\source.north) edge (H-\dest.south);

\foreach \name / \y in {1/1*\nodesep,2/2*\nodesep,3/3*\nodesep,5/5*\nodesep}
	\node[neuron, hoz,fill=red] (I2-\name) at (8*\layersep+\modelsep,-\y) {};
\node[hoz] (I2-4) at (8*\layersep+\modelsep,-4*\nodesep) {$\dots$};

\foreach \name / \y in {1/1*\nodesep,2/2*\nodesep,3/3*\nodesep,6/6*\nodesep}
	\path node [neuron, hoz,fill=red] (H2-\name) at (9*\layersep+\modelsep,-\y+0.5*\nodesep) {};
\node[hoz] (H2-4) at (9*\layersep+\modelsep,-4*\nodesep) {$\dots$};

    \foreach \source in {1,2,3,5}
        \foreach \dest in {1,2,3,6}
            \path (I2-\source.north) edge (H2-\dest.south);

\end{tikzpicture}
\caption{
Architecture of the classifier and inversion model. The classifier $F_w$ takes data $\vec{x}$ as input and produces a prediction $F_w(\vec{x})$. Such prediction vector is truncated (if necessary) to a new vector $\mathsf{trunc}(F_w(\vec{x}))$. The inversion model $G_\theta$ takes the truncated prediction as input and outputs a reconstructed data $G_\theta(\mathsf{trunc}(F_w(\vec{x})))$.
}
\label{fig:active_approach}
\end{figure*}

We propose a truncation method of training $G_\theta$ to make it aligned to the partial prediction results.
Figure~\ref{fig:active_approach} shows the architecture of the classifier $F_w$ and its inversion model $G_\theta$.
The idea is to truncate the classifier's predictions on auxiliary samples to the same dimension of the partial prediction vector on victim user's data, and use them as input features to train the inversion model, such that it is forced to maximally reconstruct input data from the truncated predictions.
Formally, let $\vec{a}$ be a sample drawn from $p_a$, and $F_w(\vec{a})$ be the classifier's prediction. 
Let $m$ be the dimension of the partial prediction vectors on victim's data.
We truncate the prediction vector $F_w(\vec{a})$ to $m$ dimensions (i.e., preserving the top $m$ scores but setting the rest to 0).
The inversion model $G_\theta$ is trained to minimize the following objective.
\begin{equation}
\mathcal{C}(G_\theta)=\mathbb{E}_{\vec{a}\sim p_a}[\mathcal{R}(G_\theta(\mathsf{trunc}_m(F_w(\vec{a}))), \vec{a})]
\end{equation}
where $\mathcal{R}$ is the loss function and we use L2 norm as $\mathcal{R}$ in this paper. Note that, it is also possible to add other loss functions such as adversarial loss in function~\ref{adv_loss} and TV loss in function~\ref{tv_loss} to make the generated data more realistic and natural. We leave it as future work.
The truncation process can be understood as a similar way of feature selection~\cite{feature_selection} by removing unimportant classes in $F_w(\vec{a})$ (i.e., those with small confidence). It helps to reduce overfitting of $G_\theta$ such that it can still reconstruct the input data from the preserved important classes.

After $G_\theta$ is trained, the adversary can feed a truncated prediction $F_w(\vec{x})$ to $G_\theta$ and obtain the reconstructed $\vec{x}$.

Our inversion model $G_\theta$ can be adopted to perform training class inference (Scenario 2), which is the same adversarial goal in MIA~\cite{model_inversion}. Training class inference can be viewed as setting $m=1$, which means the adversary knows only the class information and targets at generating a representative sample of each training class.
MIA assumes that the adversary has a white-box access 
to $F_w$ in the inference phase. Our method, on the contrary, works with a black-box access to $F_w$. 
Besides, our method can even work in the case that $F_w$ releases only the largest predicted class with confidence value, because we can approximate the total number of training classes by collecting distinct classes from the classifier's predictions on our auxiliary set. This greatly reduces the requirement of the adversary's capability to infer training classes.

Our experimental results show that the truncation method of training the inversion model improves the reconstruction quality from truncated prediction. Previous training-based approach that directly inputs the partial prediction to the inversion model produces meaningless reconstruction results (see more evaluation details in Section~\ref{sec:effect_of_adaptive_construction}).

\subsection{Joint Training of Classifier and Inversion Model}
\label{sec:joint_train}

When the adversary is the developer of the classifier, he could alternatively jointly train the inversion model with the classifier, which leads to better inversion quality. 
In particular, let $D$ be the classifier's training data.
We regularize the classification loss $L_{D}(F_w)$ (Equation~\ref{classification_loss}) with an additional reconstruction loss $R_{D}(F_w,G_\theta)$.
Intuitively, this encourages the classifier's predictions to also preserve essential information of input data in the latent space, such that the inversion model can well decode it to recover input data. 
In this paper, we use L2 norm as the reconstruction loss $R_{D}(F_w,G_\theta)$.
\begin{equation}
\label{inversion_loss}
R_{D}(F_w, G_\theta) = \frac{1}{|D|} \sum_{\vec{x}\in D} || (G_\theta(\mathsf{truc}(F_w(\vec{x})))-\vec{x}||_2^2
\end{equation}

The joint training ensures that $F_w$ gets updated to fit the classification task, and meanwhile, $G_\theta$ is optimized to reconstruct data from truncated prediction vectors.

It is worth noting that in training $G_\theta$, we find that directly using prediction vector $F_w(\vec{x})$ does not produce optimal inversion results. This is because the logits $\vec{z}$ at the output layer are scaled in $[0,1]$ (and sum up to 1) to form prediction $F_w(\vec{x})$ by the $\mathsf{softmax}$ function (see Section~\ref{sec:ml_background}), which actually weakens the activations of the output layer and thus loses some information for later decoding. To address this issue, we rescale the prediction $F_w(\vec{x})$ to its corresponding logits $\vec{z}$ in the following and use these $\vec{z}$ to replace the prediction $F_w(\vec{x})$ in our experiments.
\begin{equation}
\label{reverse_softmax}
\vec{z}=\log(F_w(\vec{x}))+c
\end{equation}
where $c$ is a constant and is also optimized during training the inversion model.

\section{Experiments}

In this section, we evaluate the inversion performance of our method and compare it with existing work.
We first introduce the experimental setup. We then evaluate three factors: the choice of auxiliary set, the truncation method and the full prediction size. We next evaluate our method in training class inference attack. Finally, we compare the inversion performance of the inversion model $G_\theta$ trained with the blackbox classifier $F_w$ and $G_\theta$ trained jointly with $F_w$.

\subsection{Experimental Setup}

\begin{table}[t]
\scriptsize
\setlength{\tabcolsep}{2pt}
\centering
\caption{Data allocation of the classifier $F_w$ and its inversion model $G_\theta$.}
\label{tb:dataset}
\begin{tabular}{l|l|l|l}
\hline
\multicolumn{2}{c|}{Classifier $F_w$} & \multicolumn{2}{c}{Inversion Model $G_\theta$} \\
\hline
\multicolumn{1}{c|}{Task} & \multicolumn{1}{c|}{Data} & \multicolumn{1}{c|}{Auxiliary Data} & \multicolumn{1}{c}{Distribution} \\
\hline
\multirow{3}{*}{FaceScrub} & 50\% train, 50\% test & FaceScrub 50\% test data & Same \\
 & 80\% train, 20\% test & CelebA & Generic \\
 & 80\% train, 20\% test & CIFAR10 & Distinct \\
 \hline
\multirow{3}{*}{MNIST} & 50\% train, 50\% test & MNIST 50\% test data & Same \\
 & 80\% train, 20\% test (5 labels) & MNIST other 5 labels & Generic \\
 & 80\% train, 20\% test & CIFAR10 & Distinct \\
 \hline
\end{tabular}
\end{table}

We perform evaluation on four benchmark image recognition datasets. For simplicity, all datasets are transformed to greyscale images with each pixel value in $[0,1]$. We detail each dataset in the following.

\begin{itemize}[leftmargin=*]
\item {\noindent \bf FaceScrub~\cite{ng2014data}.} A dataset of URLs for 100,000 images of 530 individuals. We were only able to download 48,579 images for 530 individuals because not all URLs are available during the period of writing. We extract the face of each image according to the official bounding box information. Each image is resized to $64\times 64$.

\item {\noindent \bf CelebA~\cite{liu2015faceattributes}.} A dataset with 202,599 images of 10,177 celebrities from the Internet.
We remove 296 celebrities which are included in FaceScrub and eventally we obtain 195,727 images of 9,881 celebrities.
This makes sure that our modified CelebA has no class intersection with FaceScrub.
This large-scale facial image dataset can represent the generic data distribution of human faces.
To extract the face of each image, we further crop the official align-cropped version (size $178\times218$) by width and height of 108 with upper left coordinate $(35, 70)$. Each image is resized to $64\times 64$ and also resized to $32\times 32$. 

\item {\noindent \bf CIFAR10~\cite{krizhevsky2014cifar}.} A dataset consists of 60,000 images in 10 classes (airplane, automobile, bird, cat, deer, dog, frog, horse, ship and truck). Each image is resized to $64\times 64$ and also resized to $32\times 32$.

\item {\noindent \bf MNIST~\cite{lecun1998mnist}.} A dataset composed of 70,000 handwritten digit images in 10 classes. Each image is resized to $32\times32$.
\end{itemize}

We use FaceScrub and MNIST to train the target classifier $F_w$. 
For each $F_w$ that is trained using the dataset $D$, we separately train the inversion model $G_\theta$ using the same training set $D$, training set drawn from a more generic distribution of $D$, and training set drawn from distribution that is arguably semantically different from $D$. We call these three types of training data {\em same}, {\em generic} and {\em distinct} respectively.
Table~\ref{tb:dataset} presents the data allocation for each $F_w$ and its corresponding $G_\theta$. Note that the auxiliary set of generic data distribution has no class intersection with the training data of $F_w$ (i.e., FaceScrub and MNIST).
Specifically, we have cleaned CelebA by removing the celebrities which are also included in the FaceScrub dataset. 
We randomly select 5 classes of MNIST to train $F_w$ and use the rest 5 classes to form the auxiliary set. 
The auxiliary set of distinct data distribution is composed of samples drawn from an explicitly different data distribution from the training data.

We use CNN to train $F_w$, and use transposed CNN to train $G_\theta$. The FaceScrub classifier includes 4 CNN blocks where each block consists of a convolutional layer followed by a batch normalization layer, a max-pooling layer and a $\mathsf{ReLU}$ activation layer. Two fully-connected layers are added after the CNN blocks, and $\mathsf{Softmax}$ function is added to the last layer to convert arbitrary neural signals into a vector of real values in $[0,1]$ that sum up to 1. The FaceScrub inversion model consists of 5 transposed CNN blocks. The first 4 blocks each has a transposed convolutional layer succeeded by a batch normalization layer and a $\mathsf{Tanh}$ activation function. The last block has a transposed convolutional layer followed by a $\mathsf{Sigmoid}$ activation function which converts neural signals into real values in $[0,1]$ which is the same range of auxiliary data.
The MNIST classifier and inversion model have similar architecture as FaceScrub but with 3 CNN blocks in classifier and 4 transposed CNN blocks in inversion model.
More details of model architectures are presented in Appendix~\ref{app:model}.
The FaceScrub classifier and MNIST classifier achieve 85.7\% and 99.6\% accuracy on their test set respectively (80\% train, 20\% test), which are comparable to the state-of-the-art classification performance.

\subsection{Effect of Auxiliary Set}
\label{sec:effect_of_aux}

\begin{figure*}[t]
\begin{center}
\includegraphics[width=\linewidth]{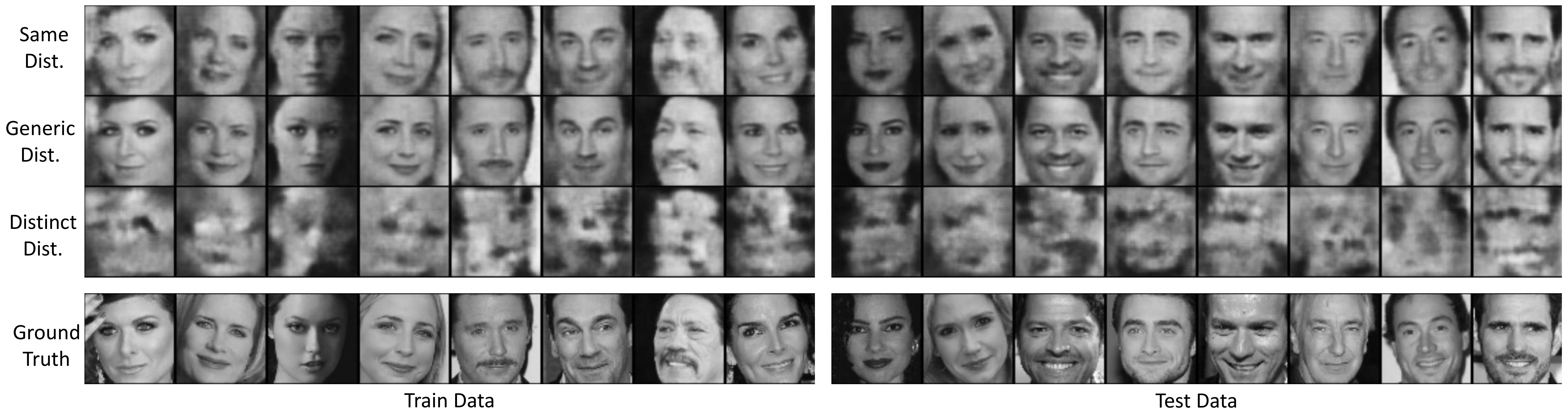}
\caption{Effect of the auxiliary set on the inversion quality. We use auxiliary set of the same (1st row), generic (2nd row) and distinct (3rd row) data distributions to train the inversion model. We perform inversion against the FaceScrub classifier $F_w$ on the $F_w$'s training data and test data.}
\label{fig:effect_of_aux_set_facescrub_large}
\end{center}
\end{figure*}

\begin{figure*}[t]
\begin{center}
\includegraphics[width=\linewidth]{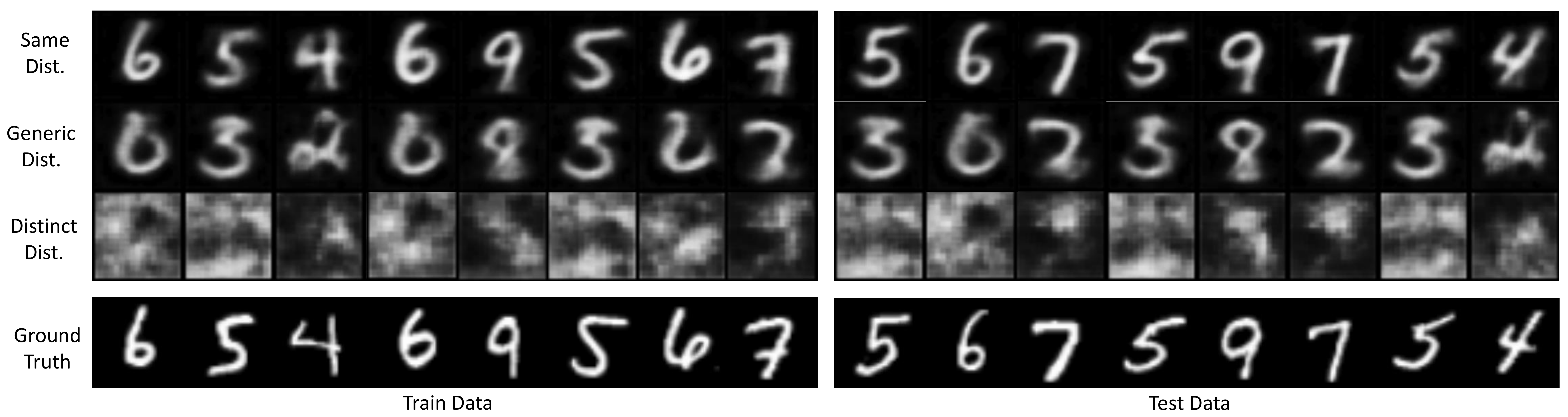}
\caption{Effect of the auxiliary set on the inversion quality. We use auxiliary set of the same (1st row), generic (2nd row) and distinct (3rd row) data distributions to train the inversion model. We perform inversion against the MNIST classifier $F_w$ on the $F_w$'s training data and test data.}
\label{fig:effect_of_aux_set_mnist_large}
\end{center}
\end{figure*}

The auxiliary set is an important factor that decides the inversion quality of $G_\theta$. 
We evaluate our inversion model $G_\theta$ by training it on different auxiliary sets (i.e., same, generic and distinct) with the blackbox $F_w$.
We use $G_\theta$ to perform data reconstruction of $F_w$'s training data and test data (Scenario 1). 
Note that in this experiment, we assume that the adversary is given full prediction values.

We present the inversion results on randomly-chosen training data and test data of the $F_w$ in Figure~\ref{fig:effect_of_aux_set_facescrub_large} and~\ref{fig:effect_of_aux_set_mnist_large}. It is clear to see that the closer the auxiliary set's data distribution is to the training data distribution, the better the inversion quality is. It is worth noting that the auxiliary set of the generic data distribution has no class intersection with the classifier's training data, which means the inversion model has never seen the training classes during its construction, but it can still accurately reconstruct the data from these classes. For example, digit 5,6,7 and 9 in MNIST are not included in the auxiliary set, yet $G_\theta$ can reconstruct these digits fairly accurately by capturing as many semantic features of these digits as possible such that they are visually like these digits. 
This further demonstrates that the generic background knowledge could be sufficient in regularizing the ill-posed inversion.
However, if the auxiliary set's data distribution is too far from the generic data distribution, the inversion is not that satisfactory as shown in the 3rd row of Figure~\ref{fig:effect_of_aux_set_facescrub_large} and~\ref{fig:effect_of_aux_set_mnist_large}. We believe that it is because the classifier cannot fully extract the semantic features from these auxiliary samples such that its predictions on them do not provide sufficient information for the inversion model to decode.

\begin{tcolorbox}
Summary \rom{1}: Even with no full knowledge about the classifier $F_w$'s training data, accurate inversion is still possible by training $G_\theta$ using auxiliary samples drawn from a more generic distribution derived from background knowledge. 
Such auxiliary set is arguably easier to obtain.
\end{tcolorbox}

\subsection{Effect of Truncation}
\label{sec:effect_of_adaptive_construction}

\begin{figure}[t]
\begin{center}
\includegraphics[width=\linewidth]{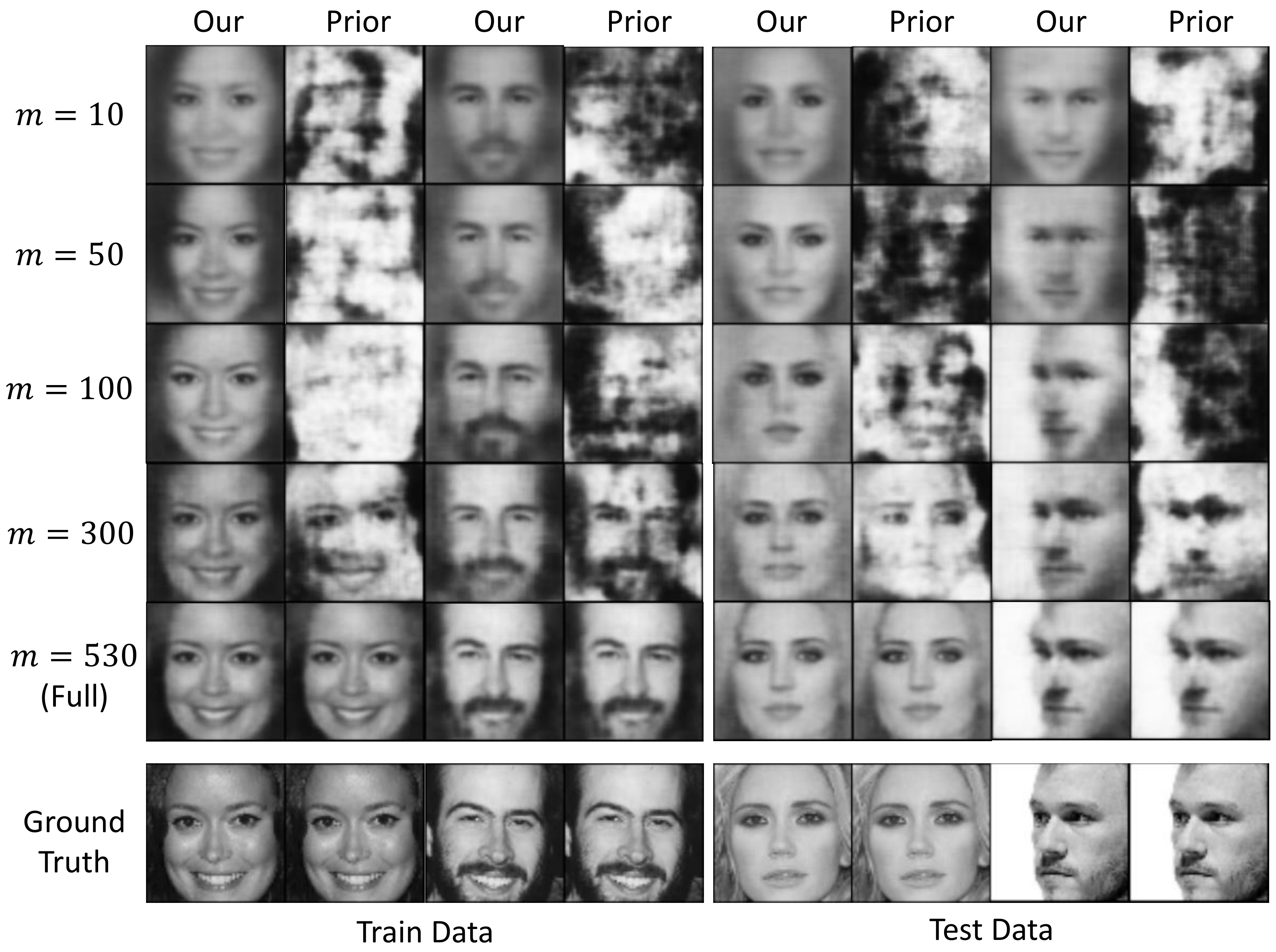}
\caption{Effect of truncation for the inversion model $G_\theta$ on the inversion quality. Inversion is performed when the adversary is given $10,50,100$ and $530$ predicted classes (denoted as $m$ in each row) from the classifier. Inversion results of our approach and previous approach are presented in odd and even columns respectively. We perform the inversion against FaceScrub classifier and present results on randomly-chosen training data and test data. The auxiliary set we use is drawn from generic data distribution.}
\label{fig:effect_of_projection_facescrub}
\end{center}
\end{figure}

\begin{figure}[t]
\begin{center}
\includegraphics[width=\linewidth]{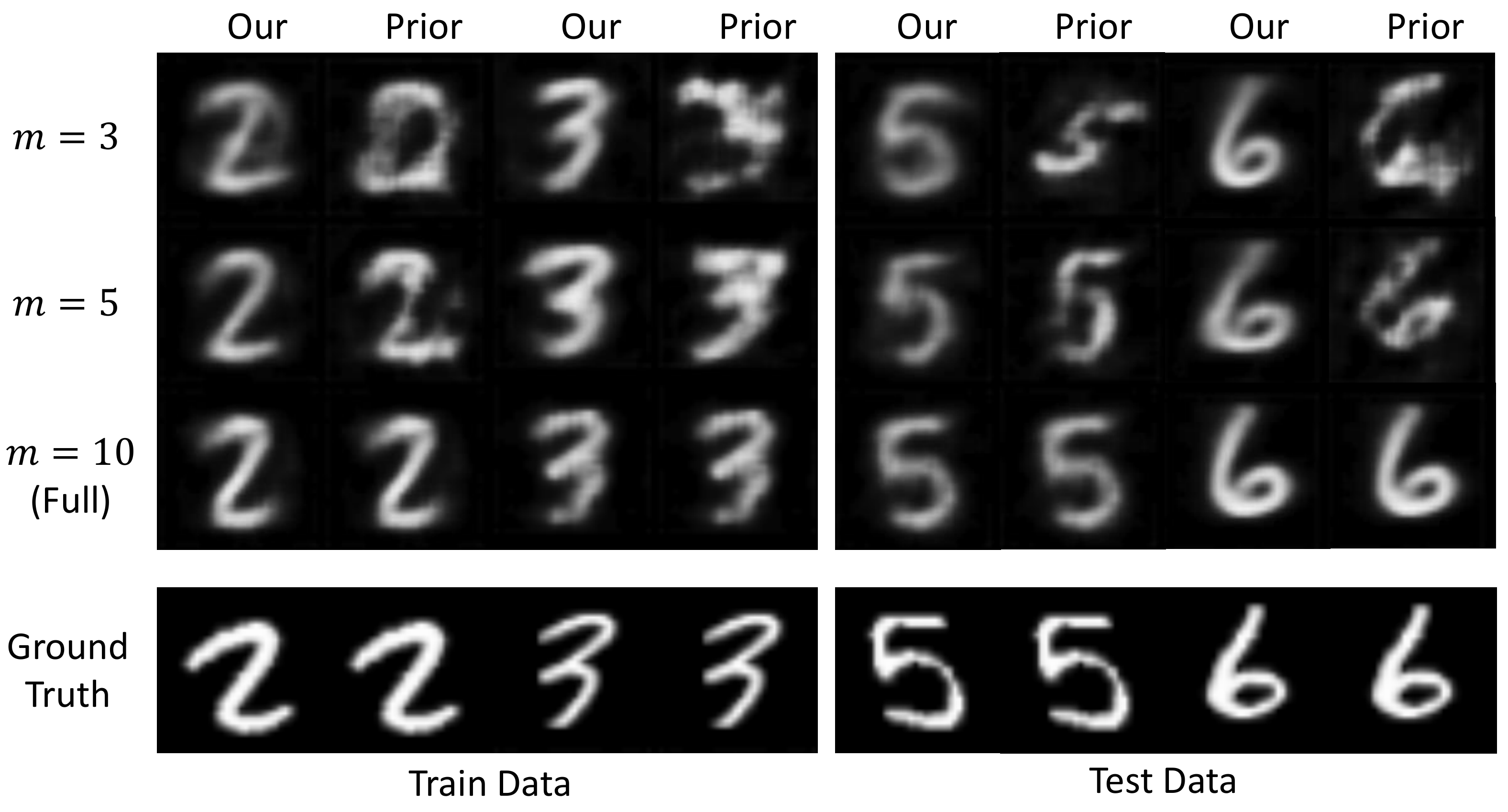}
\caption{Effect of truncation for the inversion model $G_\theta$ on the inversion quality. Inversion is performed when the adversary is given $3,5$ and $10$ predicted classes (denoted as $m$ in each row) from the classifier. Inversion results of our approach and previous approach are presented in odd and even columns respectively. We perform the inversion against MNIST classifier and present results on randomly-chosen training data and test data. The auxiliary set we use is drawn from the same data distribution.}
\label{fig:effect_of_projection_mnist}
\end{center}
\end{figure}

The truncation method of training the inversion model largely increases the inversion quality in cases where the adversary obtains only partial prediction results on victim user's data. 
Let us denote the dimension of the partial prediction that the adversary obtains on victim user's data as $m$.
We perform the inversion attack (Scenario 1) with $m=10,50,100$ and $530$ against FaceScrub classifier and with $m=3,5$ and $10$ against MNIST classifier. We present the inversion results of our truncation method and previous training-based method (i.e., without truncation) on FaceScrub classifier in Figure~\ref{fig:effect_of_projection_facescrub} and on MNIST classifier in Figure~\ref{fig:effect_of_projection_mnist}. 
The auxiliary set is composed of samples from the generic data distribution for FaceScrub classifier and from the same data distribution for MNIST classifier.
The presented training data and test data are randomly chosen.

We can see that our approach outperforms previous approach especially when the adversary is given partial prediction results. Our approach can produce highly recognizable images with sufficient semantic information for all $m$. 
Previous approach, although can generate recognizable images when $m$ is very large, produces meaningless result when $m$ is small.
For our approach, when $m$ is relatively small, it appears that the inversion result is more a generic face of the target person with facial details not fully represented. For example, in the 7th column of Figure~\ref{fig:effect_of_projection_facescrub} where the ground truth is a side face, our inversion result is a frontal face when $m=10$ and $50$, and becomes a side face when $m=100$. 
This result demonstrates that truncation indeed helps reduce overfitting by aligning $G_\theta$ to frontal facial images, and with smaller $m$, more generalization is observed.

\begin{tcolorbox}
Summary \rom{2}: Our truncation method of training the inversion model $G_\theta$ makes it still possible to perform accurate inversion when the adversary is given only partial prediction results.
\end{tcolorbox}

\subsection{Effect of Full Prediction Size}
\label{sec:effect_of_k}

\begin{figure}[t]
\begin{center}
\includegraphics[width=\linewidth]{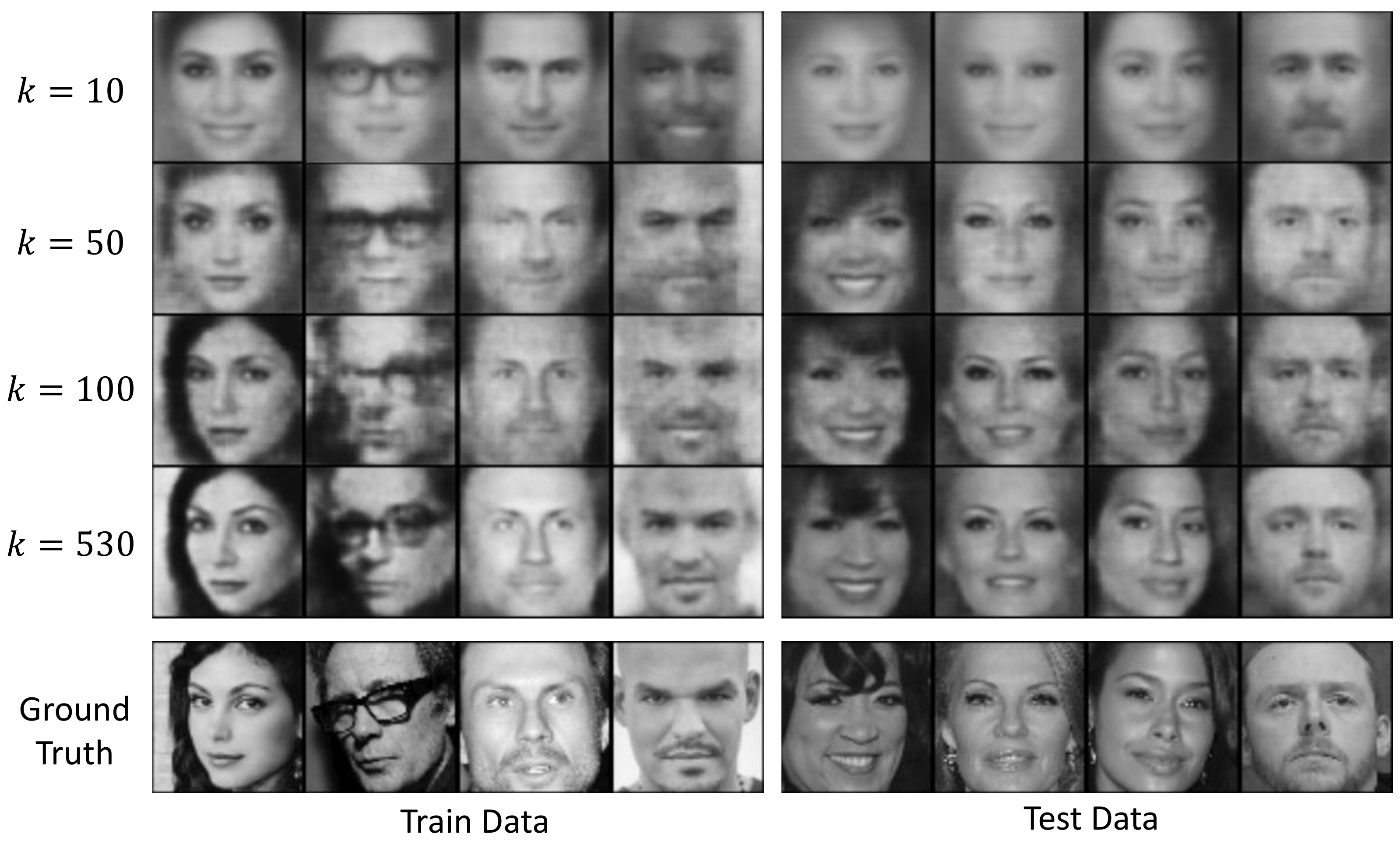}
\caption{Effect of the full prediction size on the inversion quality. Inversion is performed when the size of the full prediction vector is $10,50,100$ and $530$ classes (denoted as $k$ in each row). In each experiment, we assume the adversary obtains the full prediction vector. We perform the inversion against FaceScrub classifier and present results on randomly chosen training data and test data. The auxiliary set we use is drawn from generic data distribution.}
\label{fig:effect_of_k_facescrub}
\end{center}
\end{figure}

\begin{figure}[t]
\begin{center}
\includegraphics[width=\linewidth]{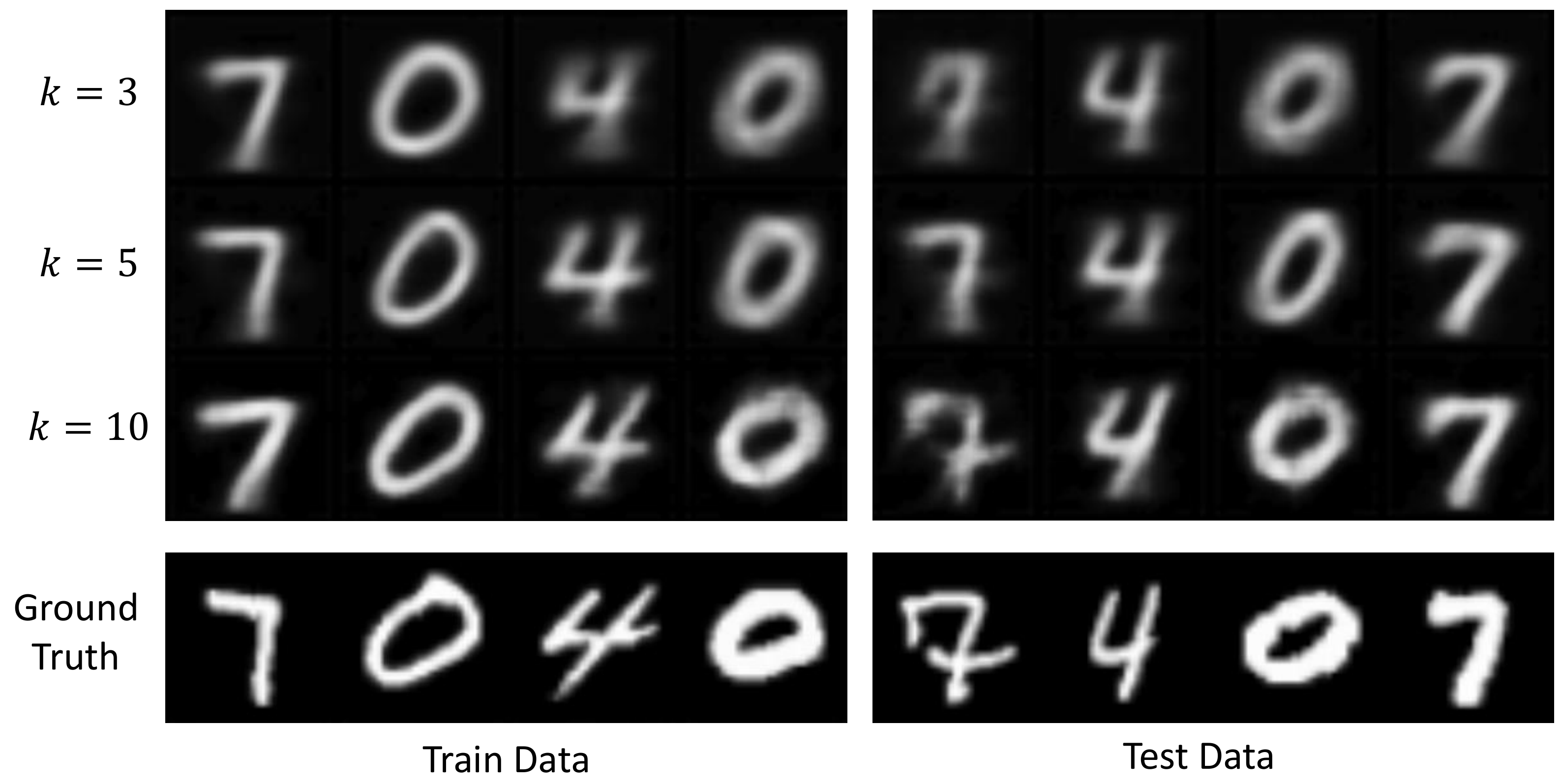}
\caption{Effect of the full prediction size on the inversion quality. Inversion is performed when the size of the full prediction vector is $3,5$ and $10$ classes (denoted as $k$ in each row). In each experiment, we assume the adversary obtains the full prediction vector. We perform the inversion against MNIST classifier and present results on randomly chosen training data and test data. The auxiliary set we use is drawn from the same data distribution.}
\label{fig:effect_of_k_mnist}
\end{center}
\end{figure}

Section~\ref{sec:effect_of_adaptive_construction} investigates the effect of $m$, the size of the truncated prediction. In this section, we investigate the effect of the size of the full prediction. Let $k$ be the dimension of the full prediction.
In our experiments, we randomly select $k$ classes from the FaceScrub and MNIST datasets as the training data of $F_w$. We set $k=10,50,100$ and $530$ for FaceScrub and $k=3,5$ and $10$ for MNIST. We present the inversion results against FaceScrub classifier in Figure~\ref{fig:effect_of_k_facescrub} and against MNIST classifier in Figure~\ref{fig:effect_of_k_mnist}. 
In all experiments, we assume that the adversary obtains the full prediction vectors (i.e., $m=k$).
The auxiliary set is composed of samples from the generic data distribution for FaceScrub classifier and from the same data distribution for MNIST classifier.
The presented training data and test data are randomly chosen.

Our experimental results show that the effect of $k$ is similar with the effect of $m$ that we have discussed. A larger $k$ leads to a more accurate reconstruction of the target data and a small $k$ leads to a semantically generic inversion of the corresponding class. This is because both $k$ and $m$ affect the amount of predicted information that the adversary can get. The factor $k$ decides the size of the full prediction vector. The factor $m$ decides how many predicted classes that the classifier releases from the $k$-dimension prediction vector.

\subsection{Training Class Inference}
\label{sec:train_class_inference}

\begin{figure}[t]
\begin{center}
\includegraphics[width=\linewidth]{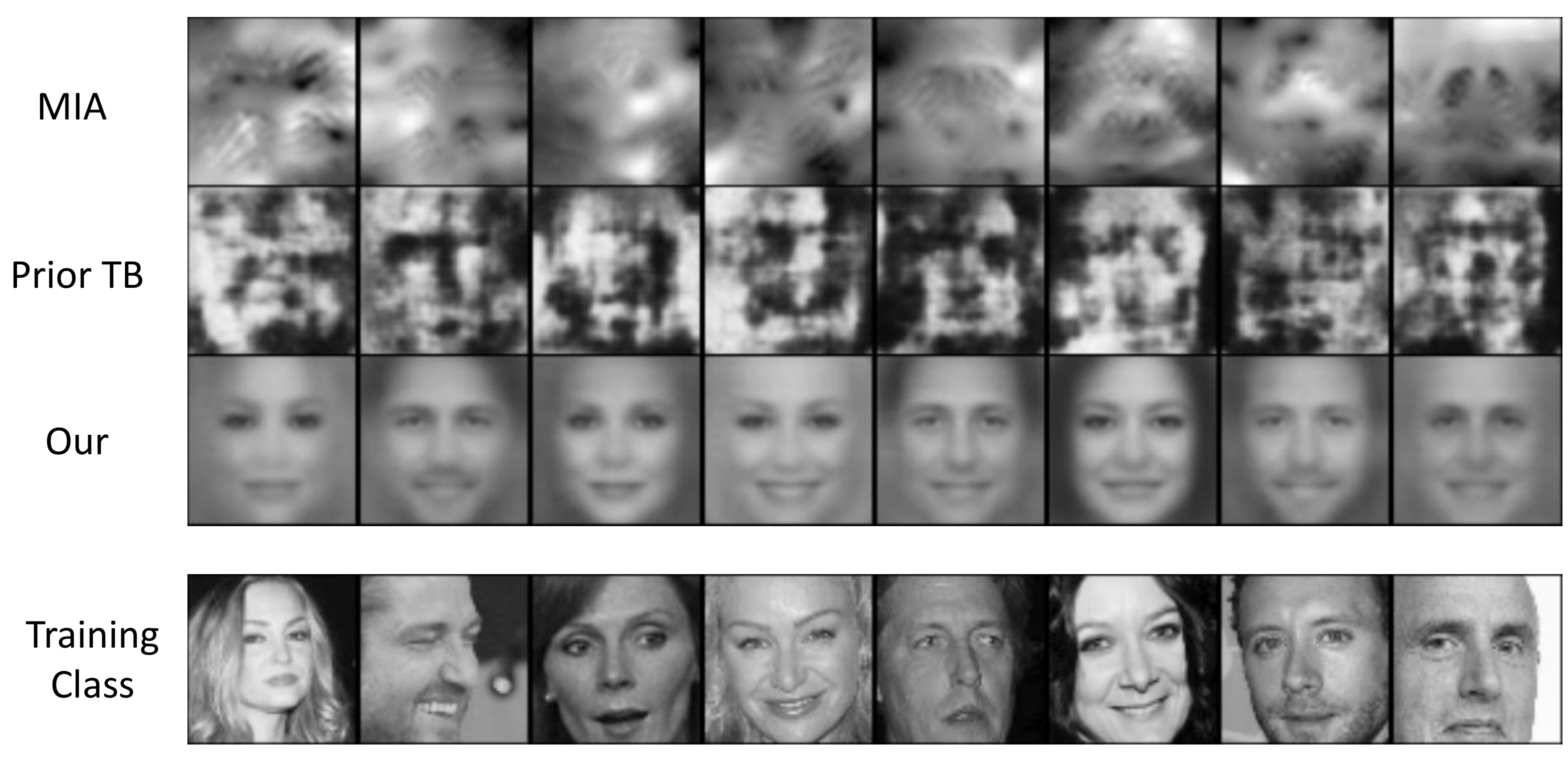}
\caption{Training class inference results. 
We present results of MIA (1st row), previous training-based (TB) inversion (2nd row) and our inversion (3rd row).
Inversion is performed against FaceScrub classifier. We use auxiliary set of generic data distribution in our inversion.}
\label{fig:train_class_inference}
\end{center}
\end{figure}

This section evaluates our inversion model $G_\theta$ on training class inference attack (Scenario 2). 
We compare our method, MIA and a previous training-based approach (which has no truncation) against the same $F_w$ trained on FaceScrub. 
For our attack, we use CelebA as the auxiliary set. It has no class intersection with FaceScrub. Hence, the inversion model $G_\theta$ trained on it has never seen the target training classes during its construction.
In particular, we first use the auxiliary set to approximate the total number of training classes, and we obtain 530 classes which is actually the true number. 
During the training of the inversion model, 
we encode the classifier's prediction result on each auxiliary sample (i.e., the largest predicted class with confidence) to a 530-dimension vector by filling rest classes with zeros. We use the encoded predictions as features to train $G_\theta$.
After $G_\theta$ is trained,
we convert each training class to a one-hot vector of 530 dimensions and feed the one-hot vectors to $G_\theta$.
The output of $G_\theta$ is the inferred image of the training class.

We present the inference results in Figure~\ref{fig:train_class_inference}. 
We can see that the inferred images by MIA are semantically meaningless and do not even form recognizable faces. This is consistent with conclusions in previous research work~\cite{ccs_gan, membership} where similar results were obtained. Our results demonstrate again that MIA is not effective when dealing with complicated network structures (i.e., CNN in our experiments). 
Previous training-based inversion also fail to produce recognizable facial images. 
Our method is able to generate highly human-recognizable faces for each person (class) by capturing semantic features such as eyes, nose and beard. An interesting find is that our method consistently produces frontal faces even though the training facial images of $F_w$ have various angles, expressions, background and even lightning. 
This demonstrates that the inversion model is well aligned to frontal faces and can capture the semantics of the training person's faces such that it can accurately infer the training classes.

\begin{tcolorbox}
Summary \rom{3}: Our method outperforms previous methods in training class inference against complex neural networks. Our experimental results show that our method can generate highly recognizable and representative sample for training classes.
\end{tcolorbox}

\subsection{Inversion Model Construction: Trained with Blackbox $F_w$ vs Trained Jointly with $F_w$}
\label{sec:different_construction_phase}

\begin{figure}[t]
\begin{center}
\includegraphics[width=\linewidth]{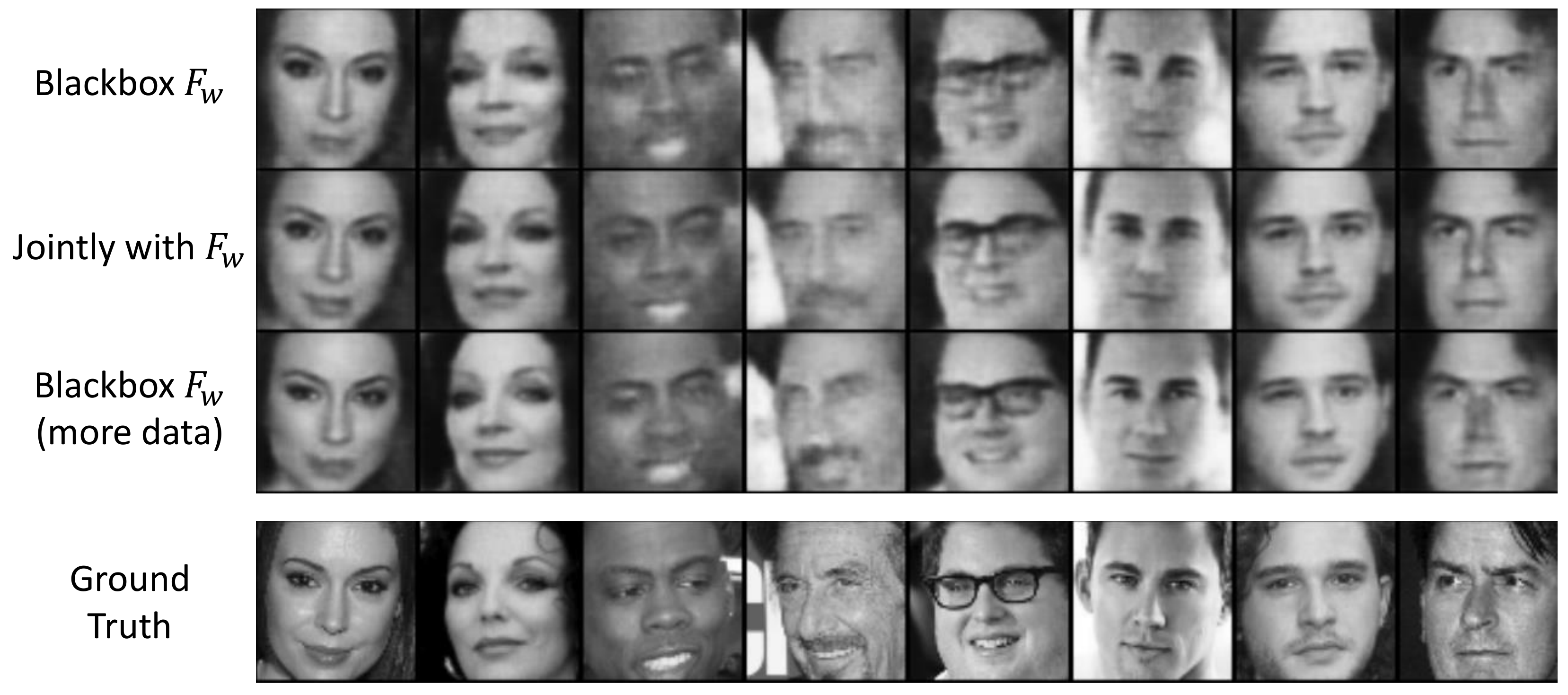}
\caption{Inversion results of the inversion model $G_\theta$: trained with blackbox $F_w$ vs trained jointly with $F_w$.
The target $F_w$ is FaceScrub.
We perform the inversion on the test data. 
For training $G_\theta$ with the blackbox $F_w$, we use the same training data of $F_w$ as the auxiliary set (1st row) and also use augmented training data with CelebA as the auxiliary set (3rd row).
All the other training details of the two construction methods are the same.}
\label{fig:different_construction_phase_facescrub}
\end{center}
\end{figure}

\begin{figure}[t]
\begin{center}
\includegraphics[width=\linewidth]{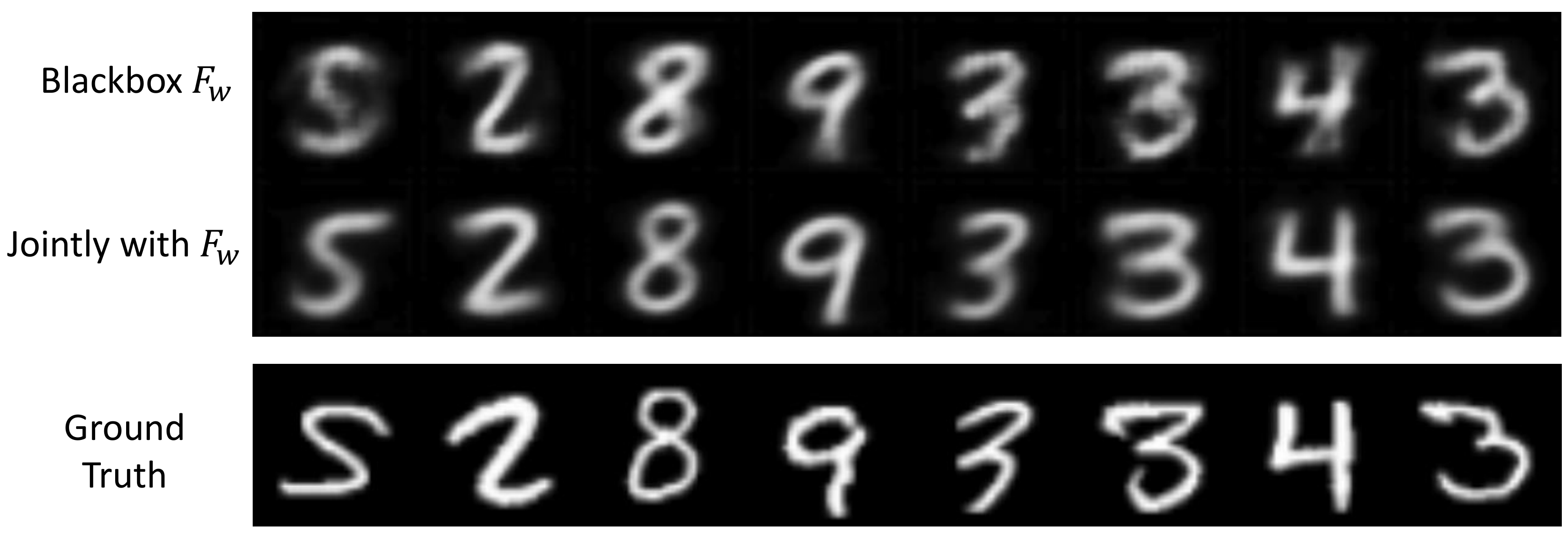}
\caption{Inversion results of the inversion model $G_\theta$: trained with blackbox $F_w$ vs trained jointly with $F_w$.
The target $F_w$ is MNIST.
We perform the inversion on the test data. 
For training $G_\theta$ with the blackbox $F_w$, we use the same training data of $F_w$ as the auxiliary set.
All the other training details of the two construction methods are the same.
}
\label{fig:different_construction_phase_mnist}
\end{center}
\end{figure}

\begin{table}[t]
\centering
\caption{Classification accuracy of the classifier $F_w$ and average reconstruction loss of the inversion model $G_\theta$. Results are reported on test set.}
\label{tb:different_construction_phase_acc}
\begin{threeparttable}
\begin{tabularx}{\columnwidth}{l|lll|lll}
\hline 
Classifier & Acc.\tnote{1} & Acc.\tnote{2} & Acc.\tnote{3} & Recon.\tnote{1} & Recon.\tnote{2} & Recon.\tnote{3}\\ \hline
FaceScrub & 78.3\% & 76.8\% & 78.3\% & 0.1072 & 0.0085 & 0.0083\\ \hline
MNIST & 99.4\% & 99.2\% & - & 0.3120 & 0.0197 & -\\ \hline
\end{tabularx}
\begin{tablenotes}
\item[1] We train $F_w$ first and then train $G_\theta$ with black-box accesses to $F_w$.
\item[2] We jointly train $F_w$ and $G_\theta$
\item[3] We train $F_w$ first and then train $G_\theta$ with black-box accesses to $F_w$. The training data is augmented with CelebA.
\end{tablenotes}
\end{threeparttable}
\end{table}

The inversion model $G_\theta$ could be trained with a fixed given blackbox $F_w$ (as in Scenario 1\&2) or trained jointly with $F_w$ (as in Scenario 3). We evaluate and compare the inversion quality of $G_\theta$ and classification accuracy of $F_w$ between the two construction methods.
To make the construction method of $G_\theta$ as the only factor that affects the inversion performance, we use the same training data of $F_w$ as the auxiliary set for the construction with blackbox $F_w$. 
All the other training details (e.g., epochs, batch size, optimizer) of the two construction methods are the same.
We use $G_\theta$ to perform test data reconstruction against the FaceScrub and MNIST classifiers. 
In particular, we split a classification dataset to 50\% as training data and the rest 50\% as test data.

We present the inversion results of the two construction methods against FaceScrub and MNIST classifiers in Figure~\ref{fig:different_construction_phase_facescrub} and~\ref{fig:different_construction_phase_mnist}.
We can see that constructing $G_\theta$ by jointly training it with $F_w$ leads to a more accurate reconstruction of the test data than that trained with the blackbox $F_w$. 
To quantify the reconstruction quality,
we present the average reconstruction loss of $G_\theta$ constructed by the two methods in the 5th and 6th columns of Table~\ref{tb:different_construction_phase_acc}. 
We can see that the joint training actually leads to a lower reconstruction loss on the test set.
This result is intuitive, because the joint training can make $F_w$'s prediction vectors also preserve essential information about the input data along with the classification information such that $G_\theta$ can well decode it for reconstruction.
However, the better reconstruction quality is achieved at the cost of lower classification accuracy. 
We present the classification accuracy in the 2nd and 3rd columns of Table~\ref{tb:different_construction_phase_acc}. 
We can see that the classification accuracy does drop because of the joint training, but it is still within an acceptable range (0.2\%-1.5\%).
Note that the classification accuracy, especially of FaceScrub (78.3\%), is a little bit lower than the accuracy (85.7\%) in earlier experiments because here we use 50\% of the original set to train $F_w$ which is less than previously 80\% of the dataset.

Certainly, the adversary (who is the malicious developer in Scenario 3) can choose to abandon the joint training. That is, $F_w$ is first trained to achieve high accuracy, and then the adversary trains the $G_\theta$. Interestingly, the adversary can choose to use two different training sets, one for $F_w$ and a different auxiliary set for $G_\theta$. In this experiment,
we augment the original training set with CelebA as the auxiliary set. We present the result in the 3rd row of Figure~\ref{fig:different_construction_phase_facescrub} and present the average reconstruction loss on the test set in the last column in Table~\ref{tb:different_construction_phase_acc}. We can see that by augmenting the training set with generic data, the inversion model $G_\theta$ trained with the blackbox $F_w$ can achieve comparable reconstruction quality to $G_\theta$ trained jointly with $F_w$.

\begin{tcolorbox}
Summary \rom{4}: Inversion model $G_\theta$ trained jointly with the classifier $F_w$ outperforms $G_\theta$ trained with blackbox $F_w$ in inversion quality but at the cost of acceptable accuracy loss. Augmenting the auxiliary set with more generic data can improve $G_\theta$ trained with the blackbox $F_w$ to be comparable to $G_\theta$ trained jointly with $F_w$, and maintain the accuracy.
\end{tcolorbox}

\section{Related Work}

While deep learning based systems are reported to attain very high accuracy in various applications~\cite{ML_Go, ML_image_accuracy}, there remains some limitations that hinder wide adoption of these systems. On the one hand, various studies have demonstrated that, similar to other data-driven applications, neural networks pose a threat to privacy~\cite{attract_info_from_ML, membership, fredrikson2014privacy,sok_ml_security}.
Besides, the fragility of neural networks is also reported in the presence of adversarial attacks~\cite{defensive_distillation_broken, NN_evasion}. On the other hand, an important drawback of neural networks is their lack of explanation capability~\cite{NN_lack_explain}. Thus, motivated either by security or interpretation reasons, many research efforts have been proposed in the literature.

\subsection{ML Privacy}

Researchers have strongly suggested that ML models suffer from various privacy threats to its training data~\cite{attract_info_from_ML, membership, fredrikson2014privacy,sok_ml_security,atrriguard}. For instance, given access to an ML model, an adversary can infer non-trivial and useful information about its training set~\cite{attract_info_from_ML}.

Shokri et al.~\cite{membership} study membership inference attacks against ML models, wherein an adversary attempts to learn if a record of his choice is part of the private training sets. 
Fredrikson et al.~\cite{model_inversion} study a model inversion attack to infer a representative sample for a target training class. 
Fredrikson et al.~\cite{fredrikson2014privacy} also propose another attack to infer sensitive attributes of training data from a released model.
Hidano et al.~\cite{PST_infer_attribute} further propose to infer sensitive attributes in a scenario where knowledge of the non-sensitive attributes is not necessarily provided.
Wu et al.~\cite{formalize_MIA} propose a methodology to formalize such attribute inference attacks.
Giuseppe et al.~\cite{ml_extraction} show that an adversary could infer the statistical information about the training dataset in the inference phase of the target model. Similarly, Hitaj et al.~\cite{ccs_gan} investigate information leakage from collaborative learning but in the training phase.
Melis et al.~\cite{inference_collaborative} investigate membership inference and property inference against collaborative ML during its training.
Wang et al.~\cite{user_level_privacy} study user-level privacy leakage against collaborative ML also during its training process.
Song et al.~\cite{ML_remember} embed sensitive information about several training data points into neural networks in its training phase by exploiting the large capacity of neural networks or massive unused capacity of model's parameter space. This enables the adversary to extract specific data points in the latter inference phase.

Some of the above mentioned attacks on training data privacy infer sensitive attributes or statistical information about the training data. Others can extract training data points but have to manipulate the training process of the model. Membership inference attack works in the inference phase but requires the data is given. Our work, on the other hands, examines how one can reconstruct either specific training data points or test data points from their prediction results in the inference phase.

\subsection{ML in Adversarial Settings}

Deep learning techniques, while achieving utmost accuracy in various application domains~\cite{ML_image_accuracy, ML_Go}, were not originally designed with built-in security. However, recent years have witnessed an ever increasing adoption of DL models for security-sensitive tasks, causing the accuracy of the models' predictions to have significant implications on the security of the host tasks.
Various works have suggested that ML models are likely vulnerable in adversarial settings~\cite{evadingML_CCS, badnets, defensive_distillation_broken}. In particular, an adversary could force a victim model to deviate from its intended task and behave erratically according to the adversary's wish. The adversary can stage these attacks by corrupting the training phase (e.g., poisoning the training set with adversarial data augmentation~\cite{poisoning_SVM, trojanning_attack_NN}, employing adversarial loss function~\cite{badnets}), maliciously modifying the victim model~\cite{defensive_distillation_broken}, or feeding the victim model with adversarially crafted samples~\cite{adversarial_examples_DL, NN_evasion}. 

In particular, Szegedy et al.~\cite{NN_evasion} leverage constrained optimization algorithm to propose an evasion technique that confuses the victim model into misclassifying an adversarial sample which is obtained by introducing a minimal and visually imperceptible perturbation to a clean sample. Papernot et al.~\cite{MLattacks_asiaccs17} further show that an adversarial sample that is crafted to target one victim model can also be effective in evading another victim model, suggesting a robustness of adversarial samples against different model configurations. While there exists various attempts to combat against adversarial samples (e.g., defensive distillation~\cite{defensive_distillation} or denoising autoencoder~\cite{denoise_adv_sample}), their effectiveness continue to be challenged by different attacks~\cite{defensive_distillation_broken}. 

On an alternative direction, Liu et al.~\cite{trojanning_attack_NN} study a trojaning attack on neural networks in which an adversary modifies a pre-trained model into a trojaned model. The trojaned model has similar performance with the pre-trained model on normal data, but will behave maliciously when presented with specific trojan trigger, which is a small stamp that is to be pasted on a clean sample. A related attack is introduced in BadNets~\cite{badnets}, wherein an adversary poisons the training set with specially crafted malicious samples so as to ``backdoor'' the output model. The backdoored model attains good accuracy on clean data, but would behave badly when presented with  specific attacker-chosen inputs. 

Our focus, on the contrary, is not to deviate ML models from their intended tasks, but rather to investigate an possibility of reconstructing input data from the model's predictions on them.

\subsection{Secure \& Privacy-Preserving ML}

In the wake of security and privacy threats posed to ML techniques, much research has been devoted to provisioning secure and privacy-preserving training of ML models~\cite{DP_DL, privacypreservingDL, obliviousML,privacy_preserving_prediction, gazelle}.
For instances, Abadi et al. study a framework to train deep learning models that enjoy differential privacy. Shokri et al.~\cite{privacypreservingDL} propose a protocol for privacy-preserving collaborative deep learning, enabling participants to jointly train a model without revealing their private training data. 
Following this work, Phong et al.~\cite{homomorphic_DL} apply homomorphic encryption to further ensure the data privacy against an honest-but-curious server.
Bonawitz et al.~\cite{secure_aggregation} present a solution for secure aggregation of high-dimensional data. In addition, systems for oblivious multi-party machine learning has also been built using trusted hardware primitive~\cite{obliviousML}. 

The threat models assumed by these techniques are to protect privacy of users' data contributed to the training set. Our work studies a different threat model wherein the adversary targets at reconstructing user data given the model's prediction results on them.

Some research work on protecting the predictions of ML models has also been proposed recently~\cite{privacy_preserving_prediction, gazelle}. For examples, Dwork and Feldman~\cite{privacy_preserving_prediction} study a method to make the model's predictions achieve differential privacy with respect to the training data.
Our work, on the other hand, studies the mapping between the prediction vectors and the input data by training another inversion model.
Juvekar et al.~\cite{gazelle} propose Gazelle as a framework for secure neural network inference in the ``machine-learning-as-a-service'' setting using cryptographic tools such that the server cannot obtain the predictions.
However, our work studies a setting where the user posts their prediction results (e.g., predicted scores of face beauty and dress sense) to social media which causes an exposure of the model's predictions.

\subsection{ML Inversion for Interpretation}

Although deep neural networks have demonstrated impressive performance in various applications, it remains not clear why they perform so well.
Much research has been working on interpreting and understanding neural networks~\cite{yosinski2015understanding, nn_explain_use_inversion, saliency_map, visualize_natural_image,nash2018inverting,interpretable_cnn}. Inverting a neural network is one important method to understand the representations learned by neural networks.

Simonyan et al~\cite{saliency_map} visualize a classification model by generating an image which maximizes the class score, and computing a class saliency map for a given image and class. Both methods are based on computing the gradient of the class score with respect to the input image.  
Zeiler and Fergus~\cite{visualize_cnn} propose the DeConvNet method to backtrack the network computations to identify parts of the image responsible for certain neural activations. They attach a deconvnet to each of the convnet layers so as to provide a propagation back to image pixels.
Du et al.~\cite{explain_use_feature_inversion} explain a DNN-based prediction by a guided feature inversion framework to identify the contribution of each feature in the input as well as to inspect the information employed by the DNN for the prediction task. Their experimental results suggest that the higher layers of DNN do capture the high-level features of the input.
Jacobsen et al.~\cite{irevnet} propose an invertible neural network that preserves information about the input features in each of their intermediate representations at each layer. They replace the non-invertible components of the RevNet~\cite{Gomez2017} by invertible ones. 
Gillbert et al.~\cite{understand_invertibility} focus on a theoretical explanation of the invertibility of CNN and propose a mathematical model to recover images from their sparse representations. However, they make several model assumptions for mathematical analysis, and there is still a gap between the mathematical model and the CNN used in practical scenarios. 
Mash et al.~\cite{nash2018inverting} train generative inversion models to interpret features learned by neural networks. They make use of autoregressive neural density models to express a distribution over input features given an intermediate representation.

The above mentioned research work inverts neural networks in order to understand and interpret them. Thus, the inversion can leverage all possible information of the model and the training data. Our work, on the contrary, inverts a neural network in the adversarial settings where the adversary's capability is limited.

\section{Conclusion}

We proposed an effective model inversion approach in the adversary setting based on training an inversion model that acts as an inverse of the original model.
We observed that even with no full knowledge about the original training data, an accurate inversion is still possible by training the inversion model on auxiliary samples drawn from a more generic data distribution.
We proposed a truncation method of training the inversion model by using truncated predictions as input to it. The truncation helps align the inversion model to the partial predictions that the adversary might obtain on victim user's data.
Our experimental results show that our approach can achieve accurate inversion in adversarial settings and outperform previous approaches.

The seemingly coarse information carried by the prediction results might lead to causal sharing of such information by users and developers.
Our work, on the other hand, shows the surprising reconstruction accuracy of the inversion model when applied in adversarial settings. 
It is interesting in the future work to investigate how other loss functions, and other generative techniques such as Generative Adversarial Network can be incorporated in the adversarial inversion problem.

\clearpage

\clearpage
\appendix

\subsection{Model Architecture}
\label{app:model}

\begin{figure}[h]
\begin{center}
\includegraphics[width=\linewidth]{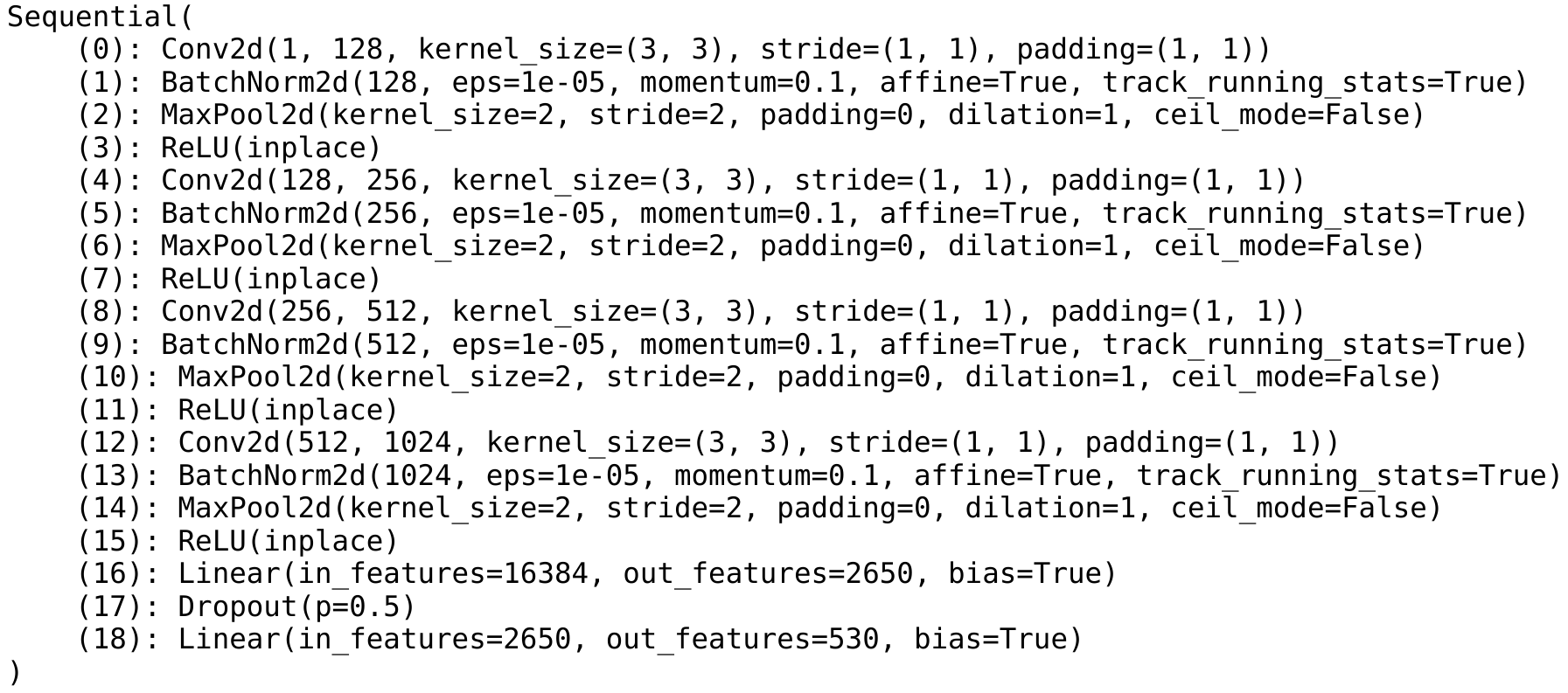}
\caption{FaceScrub classifier architecture.}
\label{fig:facescrub_classifier}
\end{center}
\end{figure}

\begin{figure}[h]
\begin{center}
\includegraphics[width=\linewidth]{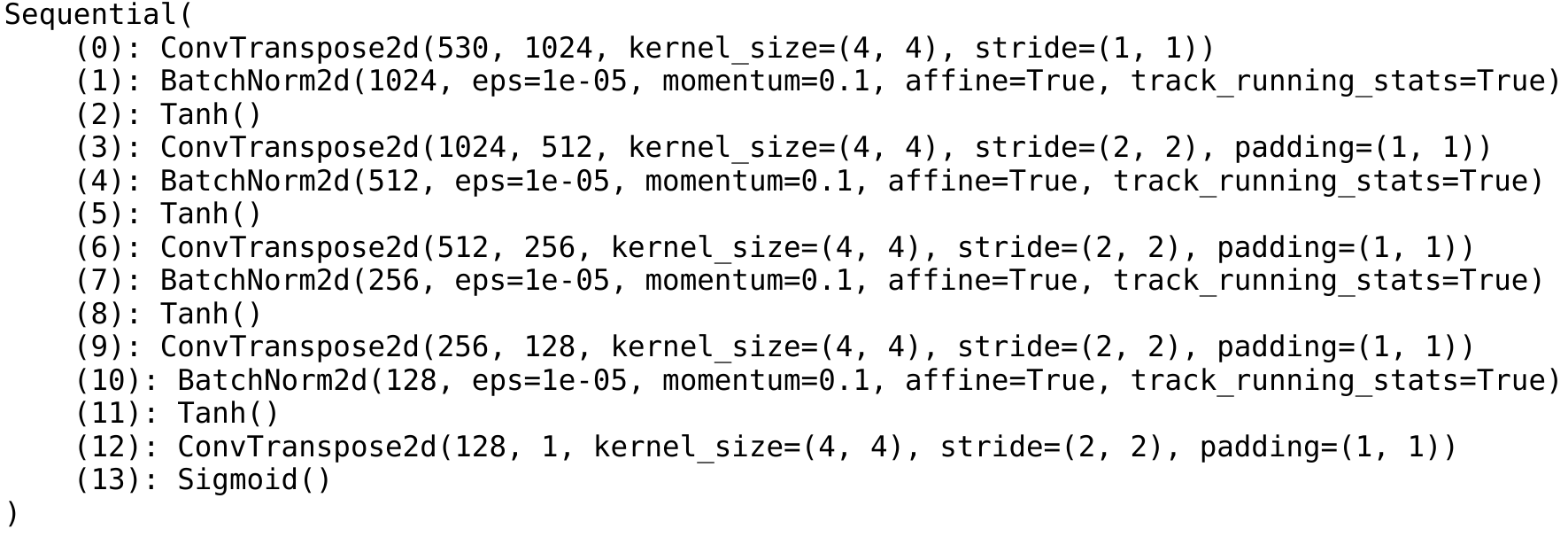}
\caption{FaceScrub inversion model architecture.}
\label{fig:facescrub_inverse}
\end{center}
\end{figure}

\begin{figure}[h]
\begin{center}
\includegraphics[width=\linewidth]{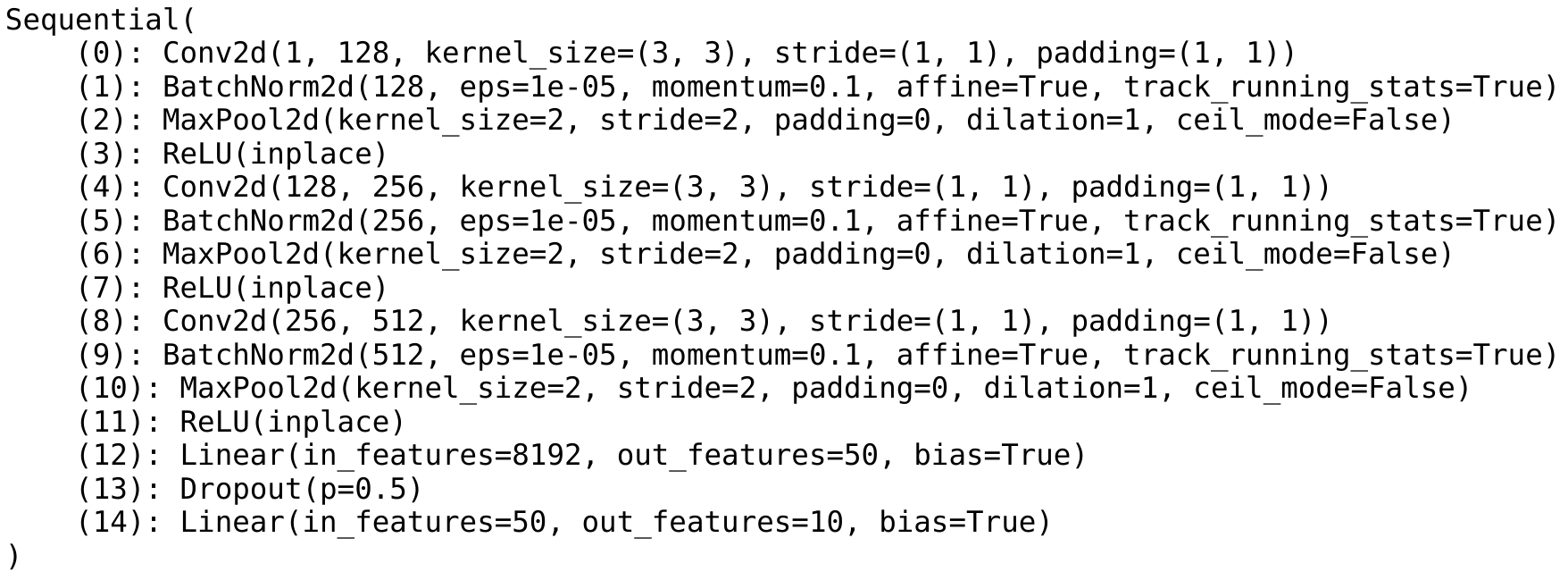}
\caption{MNIST classifier architecture.}
\label{fig:mnist_classifier}
\end{center}
\end{figure}

\begin{figure}[h]
\begin{center}
\includegraphics[width=\linewidth]{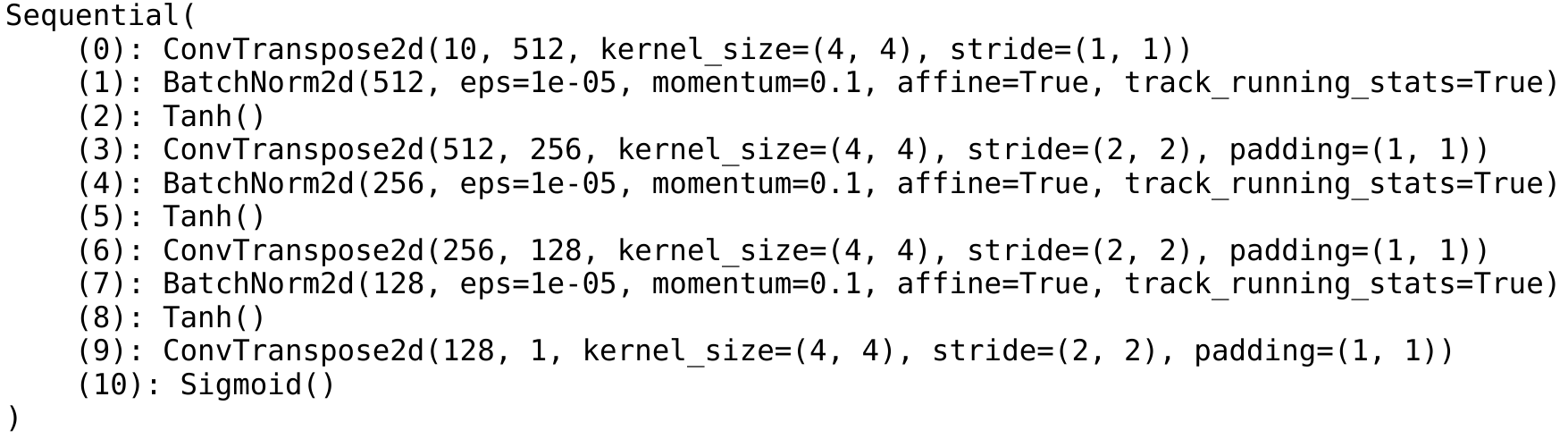}
\caption{MNIST inversion model architecture.}
\label{fig:mnist_inverse}
\end{center}
\end{figure}

\end{document}